\documentclass[%
 aip,
rsi,%
 amsmath,amssymb,
 reprint,%
]{revtex4-1}

\usepackage{graphicx}
\usepackage{dcolumn}
\usepackage{bm}
\usepackage{xcolor}

\begin{document}

\preprint{AIP/123-QED}


\title[All-dielectric metasurfaces with trapped modes]{All-dielectric metasurfaces with trapped modes: group-theoretical description}


\author{Pengchao Yu}
\affiliation{College of Physics, Jilin University, 2699 Qianjin St., Changchun 130012, China} 
\author{Anton~S.~Kupriianov}  
\affiliation{College of Physics, Jilin University, 2699 Qianjin St., Changchun 130012, China} 
\author{Victor~Dmitriev}
\affiliation{Electrical Engineering Department, Federal University of Para, PO Box 8619, Agencia UFPA, CEP 66075-900 Belem, Para, Brazil}%
\author{Vladimir~R.~Tuz}
 \email{tvr@jlu.edu.cn}
\affiliation{State Key Laboratory of Integrated Optoelectronics, College of Electronic Science and Engineering, International Center of Future Science, Jilin University, 2699 Qianjin St., Changchun 130012, China}
\affiliation{Institute of Radio Astronomy of National Academy of Sciences of Ukraine, 4 Mystetstv St., Kharkiv 61002, Ukraine}

%
\date{\today}
\begin{abstract}
An all-dielectric metasurface featuring resonant conditions of the trapped mode excitation is considered. It is composed of a lattice of subwavelength particles which are made of a high-refractive-index dielectric material structured in the form of disks. Each particle within the lattice behaves as an individual dielectric resonator supporting a set of electric and magnetic (Mie-type) modes. In order to access a trapped mode (which is the TE$_{01\delta}$ mode of the resonator), a round eccentric penetrating hole is made in the disk. In the lattice, the disks  are arranged into clusters (unit super-cells) consisting of four particles. Different orientations of holes in the super-cell correspond to different symmetry groups producing different electromagnetic response of the overall metasurface when it is irradiated by the linearly polarized waves with normal incidence. We perform a systematic analysis of the electromagnetic response of the metasurface as well as conditions of the trapped mode excitation involving the group-theoretical description, representation theory and microwave circuit theory. Both polarization-sensitive and polarization-insensitive arrangements of particles and conditions for dynamic ferromagnetic and   antiferromagnetic order are derived. Finally, we observe the trapped mode manifestation in the microwave experiment.
\end{abstract}



\maketitle
\section{Introduction}

All-dielectric resonant nanophotonics\cite{kruk_acsphotonics_2017} offers a variety of intriguing optical phenomena and enables promising practical applications. It relies on all-dielectric metasurfaces composed of \textit{subwavelength} particles. The particles are dielectric bodies which typically have a simple geometric shape (e.g. spheres, disks, cubes, parallelepipeds).\cite{Zhao_MaterToday_2009, Jahani_Nanotechnol_2016, Limonov_NatPhoton_2017} They are made of high-refractive-index materials\cite{Baranov_Optica_2017} and arranged into a lattice, wherein each particle behaves as an individual resonator sustaining a set of electric and magnetic multipole (Mie-type) modes. These modes couple to the field of incoming radiation producing a strong resonant response of the overall metasurface. 

The physics of all-dielectric resonant nanophotonics can be broadly characterized by two major phenomena: (i) sharp spatial location of electric and magnetic fields at the nanoscale, which enhances a number of nonlinear effects, such as the harmonic generation,\cite{Yang_NanoLett_2015, Shcherbakov_acsphotonics_2015, Liu_NanoLett_2016} all-optical switching,\cite{Shcherbakov_Nanolett_2015} and (ii) the multimodal interference via hybrid modes excitation. Many novel effects were discovered, such as directional light scattering,\cite{Fu_NatCommun_2013} optical magnetism,\cite{Ginn_PhysRevLett_2012, Kuznetsov_SciRep_2012} bound states in the continuum,\cite{Miroshnichenko_PhysRevB_2018, Koshelev_PhysRevLett_2018} toroidal and nonradiating optical anapole states,\cite{Miroshnichenko_NatCommun_2015, Lukyanchuk_PhysRevA_2017, Kapitanova_AdvOptMat_2018, tuz_ACSPhotonics_2018, tuz_AdvOptMat_2019} etc. 

There is a possibility to further increase the level of complexity by combining particles into clusters possessing particular spatial symmetry (or asymmetry) in order to construct topological photonic structures based on an all-dielectric metamaterial platform.\cite{Lu_NatPhotonics_2014, Slobozhanyuk_NatPhotonics_2017}

Electromagnetic features of such metasurfaces depend on both characteristics of its unit cell (i.e. on the mode composition inherent to an individual particle) and the effect of their periodic ordering inside the lattice. If the particles are grouped into a cluster to form a unit \textit{super-cell}, an electromagnetic coupling inside such super-cell results in appearing additional hybrid modes which also influence the overall electromagnetic response of the metasurface. 

Application of the method of the electromagnetic multipole expansion allows one to separate the impact of geometrical and material parameters of the individual particles from the contribution of collective effects of their interaction (coupling) within the lattice. The idea behind the multipole expansion is the following:\cite{Kallos_PhysRevB_2012} when an electromagnetic wave is incident upon a dielectric particle and therefore \textit{displacement} (polarization) currents are excited inside its volume, the currents and the particle can be replaced by multipoles of various orders (dipole, quadrupole, etc.) that act as equivalent point sources for the scattered fields. Comparing contribution of each mode to the scattering cross section of an individual particle with the resonances in the spectra of metasurface, one can separate the resonant conditions inherent to the particle only.

Although the electromagnetic multipole expansion method lies at the heart of the description of all-dielectric metasurfaces via the Mie-type modes, a more general approach based on group theoretical methods\cite{Altman_book_1991, dmitriev_book_2002} allows one to determine some of the electromagnetic properties of a metasurface based solely upon the symmetries of its unit cell. Especially, this approach becomes very powerful for metasurfaces when a unit cell contains several particles grouped into a cluster. From the transformation properties of an electromagnetic basis under symmetries of the unit cell, it is possible to determine:\cite{Padilla_OptExpress_2007} (i) the electromagnetic modes of the particles, (ii) the form of constitutive relations, (iii) polarization sensitivity or insensitivity of the metasurface, and (iv) magneto-optical response of a metamaterial or lack thereof.

Remarkably, if metasurfaces are composed of unit cells with a broken symmetry, they can feature the excitation of distinct high-$Q$ resonances which do not exist in their symmetric counterparts. These resonances are related to the inherently nonradiating modes which becomes weakly radiative when symmetry is broken. Such modes are usually referred to the \textit{trapped} (dark) modes.\cite{Zouhdi_Advances_2003, Fedotov_PhysRevLett_2007, khardikov_JOpt_2010, jain_advoptmater_2015} In the all-dielectric metasurfaces sustaining a trapped mode, the degree of asymmetry determines the strength of electromagnetic coupling between the field of incoming radiation and the polarization currents induced by this field inside the particles. Since the electromagnetic field is strongly confined inside the particles at the trapped mode resonant conditions, a metasurface can enhance several significant effects related to absorption, optical activity and optical gain.\cite{tuz_PhysRevB_2010, Tuz_JOpt_2012, Tuz_EurPhys_2011, Khardikov2016, Tong_OptExpress_2016, Cui_acsphotonics_2018}

The all-dielectric metasurfaces sustaining a trapped mode are considered as an efficient platform for the engineering optical magnetism. It was particularly revealed\cite{Khardikov_JOpt_2012, Zhang_OptExpress_2013, Tuz_OptExpress_2018} that at the resonant frequency of the trapped mode excitation, the polarization currents induced in the dielectric particles by incident wave have a circular flow twisting around the particle center, whereas the magnetic field direction in the particle center is oriented orthogonally to the metasurface plane forming out-of-plane magnetic dipole moment. In the super-cell based metasurfaces the magnetic moments induced in a set of particles forming a cluster can be oriented either in the same direction demonstrating a dynamic ferromagnetic (FM) order or in the opposite directions resembling a dynamic antiferromagnetic (AFM) order.\cite{sayanskiy_arXiv_2018} 

In the present paper, we suggest a systematic study of an all-dielectric metasurface featuring trapped modes. The analysis is based on group-theoretical methods, representation theory and microwave circuit theory. Our goal is to reveal fundamental properties of the metasurface, such as the necessary changes in symmetry for excitation of a trapped mode, conditions for the FM and AFM orders of resonant magnetic field, appearance of the electric dipole mode as well as higher-order modes, compositions of the eigenwaves and their degeneracy, polarization dependence, cross-polarization effects and chirality. 

For the further discussion, we have chosen a specific dark mode of a cylindrical dielectric resonator and an array with a high symmetry of the unit cell, namely, which has the symmetry of a square. With small modification, one can obtain in this array several lower symmetries of the unit cell. Therefore, this scheme allows one to consider and compare properties of arrays from the point of view of their symmetry.

The rest of the paper is organized as follows: In Sec.~\ref{sec:particle} we study characteristics of an isolate dielectric particle presenting either non-perturbed or perturbed disk.  Electromagnetic features of  metasurface, whose unit super-cell is composed of four perturbed disks ordered in accordance with the groups $C_{4v}$, $C_{2v}$,  $C_{4}$, $C_{2}$ and $C_s$, are studied in Sec.~\ref{sec:metasurface}. Symmetry conditions for the FM and AFM orders are studied theoretically and confirmed experimentally in Sec.~\ref{sec:magneticorder}. The effect of electromagnetic coupling between the resonators in a lattice and polarization dependence of the resonator excitation are discussed in Sec.~\ref{sec:coupling} and Sec.~\ref{sec:polarization}, respectively. Finally, in Sec.~\ref{sec:concl} we summarize the paper. The scattering cross section of a perturbed particle is given in Appendix A for reference. 
\section{Isolated particle characterization}
\label{sec:particle}
An overall electromagnetic response of a metasurface is mainly dependent on properties of its unit cell, so it is necessary to reveal first a modal composition of a single constituent particle of a metasurface. It can be derived considering spatial in-plane symmetry of the particle. Our treatment is based upon the symmetry operations of the constituent particle about a point in space, i.e. it involves the point group theory. 

The derivation is performed under the assumption of small dimensions of the particle compared to the wavelength of the incident radiation (i.e. the particle is subwavelength one). We consider a dielectric particle of two shapes: a solid disk (non-perturbed cylindrical resonator) and a disk having a round eccentric hole (perturbed cylindrical resonator). For the latter one, we expect to get access to a trapped mode which is a nonradiating mode of the non-perturbed resonator.
\subsection{Non-perturbed dielectric resonator}
\begin{figure}[htp]
\centering
\includegraphics[width=75mm]{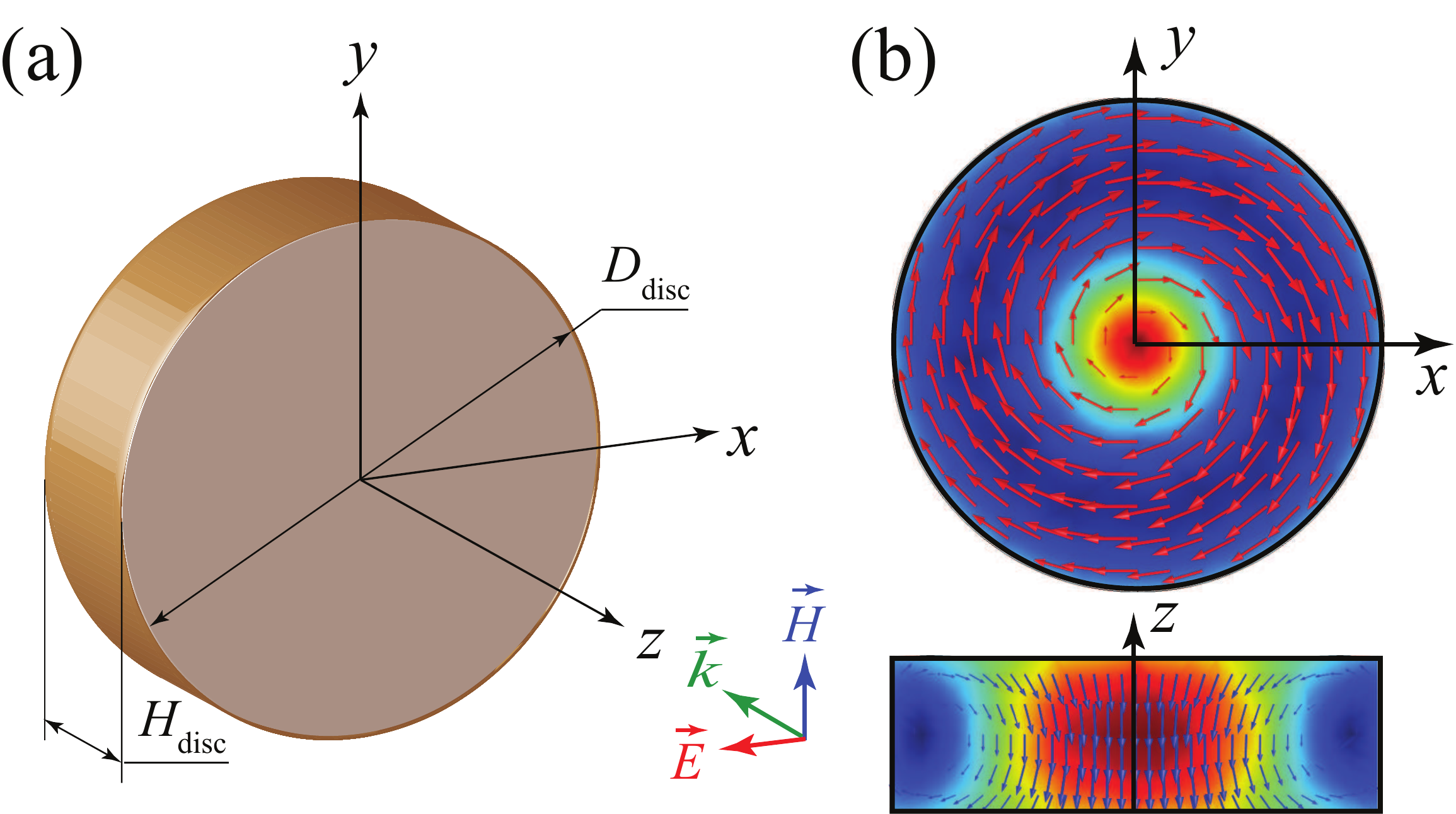}
\caption{(a) Geometry of an isolated non-perturbed dielectric resonator (disk), and (b) distributions of the displacement current (red arrows) and magnetic field (blue arrows) of the dark Mie-type mode which corresponds to the TE$_{01\delta}$ mode of the  resonator.}
\label{fig:fig1}
\end{figure}

We consider a dielectric disk excited by a plane wave with linear polarization which propagates along the cylinder's axis (${\bf k} = \{0,0,-k_z\}$, i.e. normal incidence,  Fig.~\ref{fig:fig1}(a)). The height and diameter of the disk are $H_\textrm{disk}$ and $D_\textrm{disk}$, respectively. It is made of a nonmagnetic material with permittivity $\varepsilon_d$. The disk is placed on a dielectric substrate whose thickness and permittivity are $H_\textrm{subs}$ and $\varepsilon_s$, respectively.

For the normally incident plane wave, it is sufficient to consider symmetry of the particle in the $x-y$ plane only.  The geometry of the cylindrical resonator in the $x-y$ plane is a circle. It is described by the group of symmetry $C_{\infty v}$ (in Sch\"oenflies notation\cite{dmitriev_book_2002}).
We consider the resonator placed on a dielectric substrate, therefore, the plane of symmetry perpendicular to the $z$-axis is absent. $C_{\infty v}$ is a continuous two-dimensional rotation group. It consists of any rotation around the $z$-axis and an infinite number of vertical planes of symmetry passing through this axis. 

In what follows, we shall restrict our discussion to the dark Mie-type mode which is the lowest transverse electric  mode TE$_{01\delta}$ of the cylindrical resonator. Such a mode cannot be excited in the resonator by a normally incident linearly polarized wave. The structure of the field inside the resonator is dictated by symmetry of the disk. We are interested in the electromagnetic field components $E_x$, $E_y$, and $H_z$ which characterize the TE$_{01\delta}$ mode, whereas the field outside the resonator contains other components which can be considered analogously. Indeed, the electric field $\bf E$ of this mode inside the resonator in the $x-y$ plane has a form of concentric circles (Fig.~\ref{fig:fig1}(b)), i.e. it consists of the $E_x$ and $E_y$ components. 

\begin{figure}[htp]
\centering
\includegraphics[width=75mm]{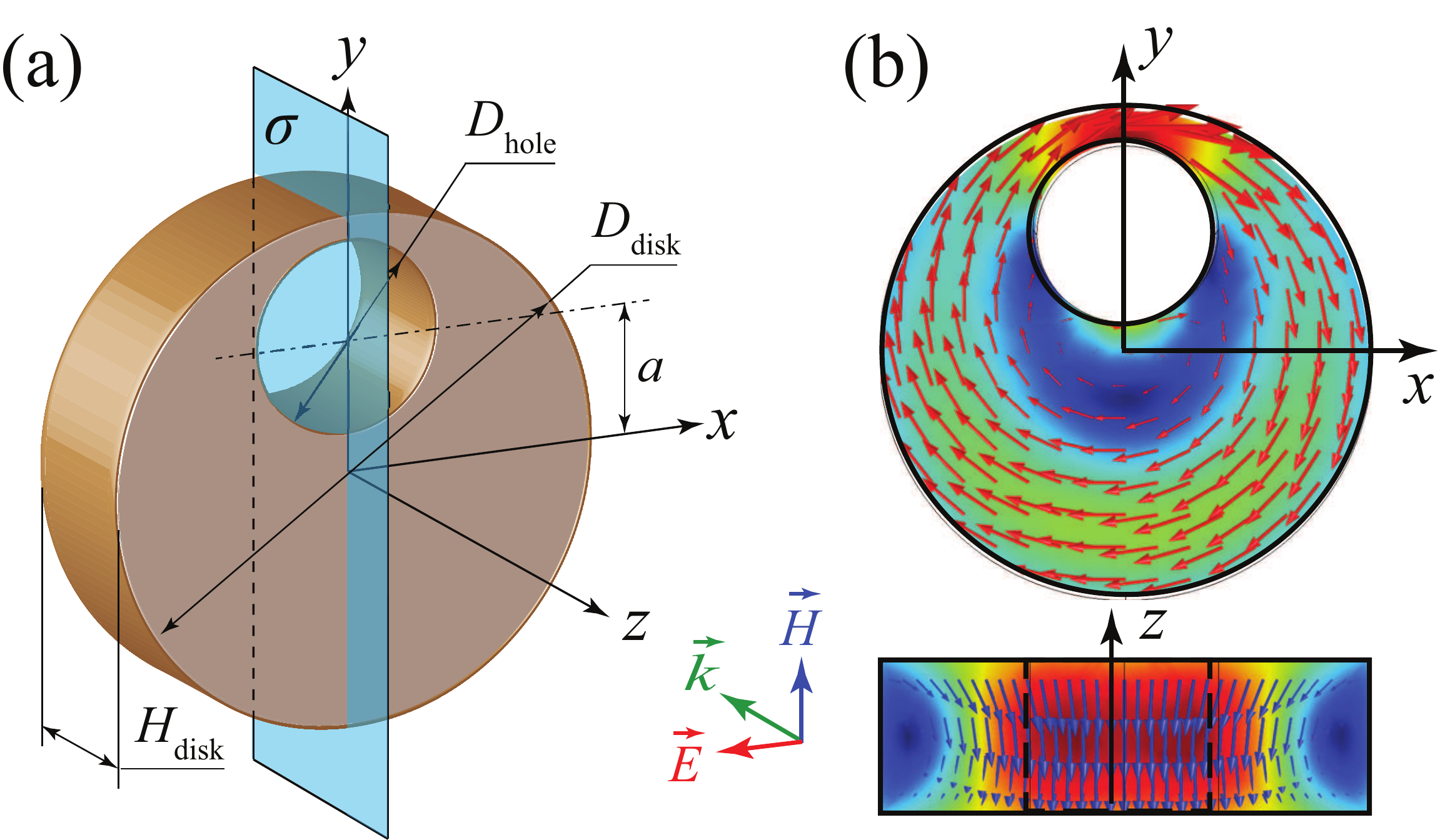}
\caption{Same as in Fig.~\ref{fig:fig1} but for an isolated perturbed dielectric resonator.}
\label{fig:fig2}
\end{figure}
\subsection{Perturbed dielectric resonator}
The cylindrical resonator under study supports the TE$_{01\delta}$ mode which does not have electromagnetic field components allowing  to couple it to the field of incident  wave. In order to access this mode, one needs to perturb the resonator, i.e. to change its symmetry providing the necessary coupling. This can be made by many ways. Here we consider a particular perturbation of the cylindrical resonator by an eccentric through hole (feeding hole) in the disk shown in Fig.~\ref{fig:fig2}(a). The hole changes the structure of the inner electromagnetic field. Nevertheless, for this case we shall continue to deal with the components $E_x$, $E_y$ and $H_z$, since these components remain to be dominant at the frequency of the TE$_{01\delta}$ mode excitation. 

The hole can be of arbitrary shape, but in order to simplify our discussion, we introduce one restriction: it must have the plane of symmetry $\sigma$ coinciding with one of the planes of the group of the non-perturbed resonator $C_{\infty v}$.

From the technological point of view, the optimal form for the feed is a round hole. The diameter of the hole is denoted as $D_\textrm{hole}$. If the hole is dislocated from the center of the disk on the distance $a$, the rotational symmetry as well as all other planes of symmetry appear to be broken, except the plane $x=0$. Therefore, the full axial symmetry of the group $C_{\infty v}$ is reduced  to the discrete group  $C_s$ of the perturbed resonator. The symmetry $C_s$ is defined by only one plane $x=0$ which is denoted here by $\sigma$. Belonging of the electromagnetic fields in the resonator to the IRREP's of this group is given in Tab.~\ref{tab:Cs1}. Thus, the group $C_s$ defines the local symmetry of the isolated perturbed resonator.

\begin{table}
	\begin{center}
		\caption{Irreducible representations 
			of the group $ C_s$ and transformation properties
			of perturbed resonator fields} 
		{
			\begin{tabular}{c@{\hspace{4.0mm}}c@{\hspace{4.0mm}}c@{\hspace{4.0mm}}c}
				\hline
				\fbox{$C_{s}$}  & $e$   & $\sigma$  & Field    \\
				\hline
				$\Gamma_1$     & 1   &  1  &           \\
				$\Gamma_2$     & 1   & $\!\!\!\!-1$  &  $E_x$, $E_y$, $H_z$ \\
				\hline
			\end{tabular}
		}
		\label{tab:Cs1}
	\end{center}
\end{table}

The level of the geometric asymmetry of the particle depends on both the parameter $a$ and the diameter of the hole $D_\textrm{hole}$ when $a\neq 0$. Such a perturbation by a hole can transform a dark (nonradiating) TE$_{01\delta}$ mode into a bright (radiating) one. Coupling between the incident wave ${\bf E}^\textrm{inc}$ (which is constant over $x-y$ plane) and the TE$_{01\delta}$ mode of the non-perturbed resonator is defined by the overlapping integral of their electric fields:
\begin{equation}
\Omega = {\bf E}^\textrm{inc} {\cdot} \int _\textrm{hole} {{(\bf E}}_{\textrm{TE}_{01\delta}})^\ast dx dy
\label{eq:overlapint}
\end{equation}

From the above equation one can see, that the coupling coefficient $\Omega$ depends on the parameters $a$ and $D_\textrm{hole}$ of the resonator. It is maximized when the electric field of the incident wave is perpendicular to the plane of symmetry $\sigma$ of the resonator. 
A simple physical interpretation of this result is as follows: (i) the electric field component of the incident wave which is perpendicular to $\sigma$, i.e. oriented along the electric field lines of the TE$_{01\delta}$ mode, excites this mode, and (ii) the component parallel to the plane $\sigma$, does not excite the TE$_{01\delta}$ mode of the resonator. Notice that the higher values of $\Omega$ lead to lower quality factor of the TE$_{01\delta}$ mode.

The perturbation which provides coupling of the incident wave and the resonator, produces also an electric dipole moment in the $x-y$ plane. This dipole moment is oriented normally to the plane of symmetry $\sigma$ (Fig.~\ref{fig:fig2}(a)). The existence of this moment is explained by the fact that due to lack of the plane of symmetry $y=0$, the displacement currents ${\bf j}_d = \partial {\bf D}/\partial t$ in the regions $y>0$ and $y<0$ of the resonator do not compensate each other any more (see also contribution of both electric and magnetic dipole moments to the scattering cross section of a perturbed particle in Appendix~A).
\section{Metasurface characterization}
\label{sec:metasurface}
Further we consider a metasurface presenting a periodic array of square unit super-cells. Each  super-cell consists of four particles  ordered according to a particular symmetry (Fig.~\ref{fig:fig3}). These particles are perturbed dielectric resonators discussed above. The metasurface is illuminated by a plane electromagnetic wave under normal incidence conditions (${\bf k} = \{0,0,-k_z\}$). The field of the incident wave produces a collective oscillation of displacement currents in the dielectric particles due to the near-field coupling between them. Our goal here is to reveal some fundamental properties of the metasurface possessing  geometrical  symmetry of its unit super-cell using group-theoretical methods. We are mainly interested in the spectral  manifestation of the dark TE$_{01\delta}$ mode which is inherent to the resonators forming the metasurface.
   
In the analyzed range of geometrical and physical parameters, the near-field characteristic is defined mostly by the local symmetry of an isolated particle and its interaction with neighbouring particles within the unit super-cell, whereas the metasurface response in the far-field depends also on the global symmetry of the array. In the long-wavelength approximation, the translational symmetry of the array can be excluded from consideration. Thus, the global symmetry of the metasurface is defined by the point symmetry of the unit super-cell, and the overall characteristic of the metasurface can be described in terms of scattering matrix.
\subsection{Possible symmetries of array unit cell with four dielectric resonators}
An arrangement of the four non-perturbed resonators (with the symmetry $C_{\infty v}$) within the unit super-cell gives the maximal symmetry group $C_{4v}$. In the arrangement of the perturbed resonators whose position within the super-cell is fixed as for the non-perturbed ones, orientation of the holes defines the symmetry of the array. Here, the maximal symmetry is also $C_{4v}$. Notice that this is a case when the symmetry of the cluster-based structure (which is $C_{4v}$) is higher than the symmetry of the individual element (which is $C_s$).

\begin{figure}[htp]
\centering
\includegraphics[width=85mm]{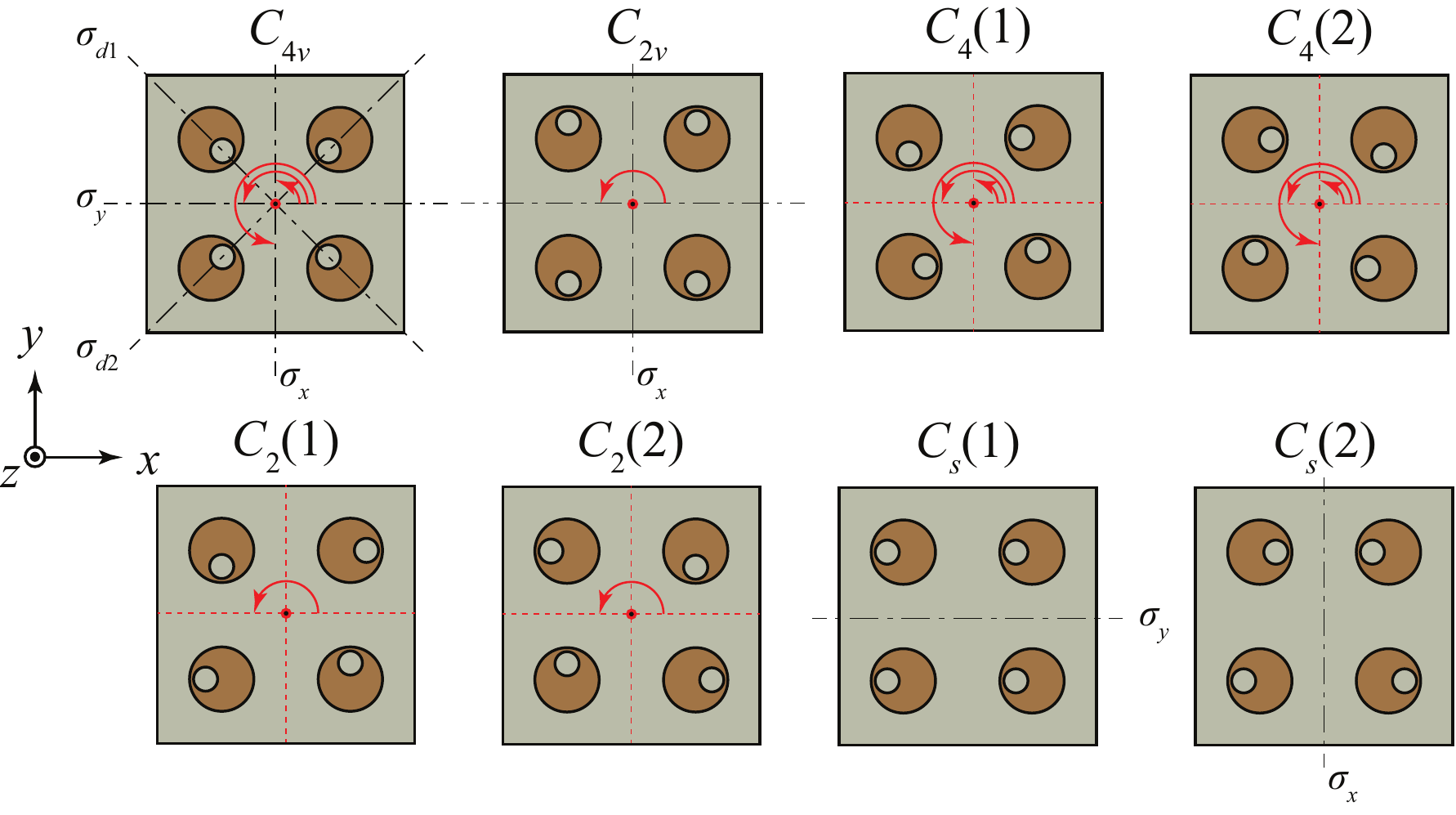}
\caption{A set of unit super-cells of the all-dielectric metasurface and corresponding groups and elements of symmetry.}
\label{fig:fig3}
\end{figure}

The symmetry, corresponding to the four-fold rotation  and  the two-fold rotation  around the $z$-axis of the unit super-cell, the plane of symmetry $\sigma_{x}$ (the plane $x=0$), the plane of symmetry $\sigma_y$ (the plane $y=0$) and the diagonal planes of symmetry $\sigma_{d1}$ and  $\sigma_{d2}$ are shown in Fig.~\ref{fig:fig3} for further reference. All the planes of symmetry pass through the $z$-axis of the unit super-cell. 

Excluding from our consideration the group $C_{1}$ which corresponds to absence of symmetry, the group $C_{4v}$ has four subgroups. They are $C_{2v}$, $C_{4}$, $C_{2}$, and $C_s$ (see the group subordination scheme (group tree) given in Fig.~2 of Ref.~\onlinecite{Dmitriev_Metamat_2011}). It is the exhaustive list of possible symmetries of a plane square. Thus, symmetry of the unit super-cell can be described by anyone of the above groups. All these symmetries can be easily obtained in the given unit super-cell by simple rotations of the individual resonators around their axes.

\begin{table*}
	\begin{center}
	\caption{Irreducible representations of the group $C_{4v}$ } 
{
\begin{tabular}{c@{\hspace{0mm}}c@{\hspace{0mm}}c@{\hspace{0mm}} 
c@{\hspace{0mm}}c@{\hspace{0mm}}c@{\hspace{0mm}}c@{\hspace{0mm}}
c@{\hspace{0mm}}c@{\hspace{0mm}}c} \hline
\fbox{$C_{4v}$}   & e   & $C_2$ & $C_4$ & $C^{-1}_4$ & $\sigma_x$ &
$\sigma_y$ & $\sigma_{d1}$ & $\sigma_{d2}$   \\  \hline
$\Gamma_1$ & 1   & 1   & 1        & 1        & 1        & 1
& 1  &  1     \\
$\Gamma_2$ & 1   & 1   & 1        & 1        & $\!\!\!\!\! -1$
&$\!\!\!\!\! -1$ &$\!\!\!\!\! -1$ &$\!\!\!\!\! -1$        \\
$\Gamma_3$ & 1   & 1   &$\!\!\!\!\! -1$       &$\!\!\!\!\! -1$
& 1        & 1           &$\!\!\!\!\! -1$ &$\!\!\!\!\! -1$      \\
$\Gamma_4$ & 1   & 1   &$\!\!\!\!\! -1$       &$\!\!\!\!\! -1$
&$\!\!\!\!\! -1$       &$\!\!\!\!\! -1$          & 1  & 1       \\
$\Gamma_5$ &
$\left( \!\!\!
\begin{array} {cc}
 1 & 0   \\
 0 & 1
\end{array} \!\!\!
\right)$ &
$\left(\!
\begin{array} {cc}
 \!\!\!\!-1 & 0   \\
 0 & \!\!\!\!-1
\end{array} \!\!\!
\right)$ &
$\left( \!\!\!
\begin{array} {cc}
 0 &\!\!\!\!-1   \\
 1 & 0
\end{array} \!\!\!
\right)$ &
$\left(\!
\begin{array} {cc}
         0 & 1   \\
\!\!\!\!-1 & 0
\end{array}\!\!\!
\right)$ &
$\left(\!\!\!
\begin{array} {cc}
 1 &          0   \\
 0 & \!\!\!\!-1
\end{array}\!\!\!
\right)$ &
$\left( \!
\begin{array} {cc}
 \!\!\!\!-1 & 0   \\
          0 & 1
\end{array}\!\!\!
\right)$ &
$\left(\!
\begin{array} {cc}
          0 & \!\!\!\!-1   \\
 \!\!\!\!-1 & 0
\end{array}\!\!\!
\right)$ &
$\left( \!\!\!
\begin{array} {cc}
 0 & 1   \\
 1 & 0
\end{array}\!\!\!
\right)$
\\ \hline
\end{tabular}
}
\label{tab:C4v}
\end{center}
\end{table*}

In what follows, we successively consider the electromagnetic characteristics of the metasurfaces, whose unit super-cell is described by the corresponding group. We distinguish polarization of the incident wave with the vector $\bf E$ oriented either along the $x$-axis ($x$-polarization) or along the $y$-axis ($y$-polarization), or along the unit super-cell diagonals ($d$-polarization). 
\subsection{Array with $C_{4v}$ symmetry}
In order to reach the highest symmetry $C_{4v}$, orientations of the planes of symmetry $\sigma$ of the perturbed resonators must be along the diagonal symmetry elements $\sigma_{d1}$ and $\sigma_{d2}$ of the unit super-cell (Fig.~\ref{fig:fig3}).

\subsubsection{Scattering matrix}

Using the microwave circuit theory,\cite{Dmitriev_IEEEAntennas_2013} we define the $2 \times 2$ scattering matrix (which, in fact, is the reflection matrix), that relates the components $E_{1}^\textrm{inc}$, $E_{2}^\textrm{inc}$ of the electric field of the incident wave with the components $E_{1}^\textrm{ref}$, $E_{2}^\textrm{ref}$ of the electric field of the reflected  wave as follows: 
\begin{equation}
\left(\!\!
\begin{array} {c}
E_{1}^\textrm{ref}  \\[1mm]
E_{2}^\textrm{ref}
\end{array} \!\!
\right)=
{\bf {\bar S}}
\left( \!\!
\begin{array} {c}
E_{1}^\textrm{inc}  \\[1mm]
E_{2}^\textrm{inc}
\end{array} \!\!
\right),
\label{eq:2.7.5}
\end{equation}
where port 1 and port 2 are oriented in the $x$-direction and $y$-direction, respectively. Analogously one can introduce the transmission matrix. We shall restrict ourselves by consideration of the reflection matrix only, because the symmetry properties of the reflection and transmission  matrices are the same. The eigenvalue problem for the scattering matrix 
is $\bf {\bar S}{\bf E}=s{\bf E}$, where $\bf E$ is an eigenvector and $s$ is the corresponding eigenvalue.

Using the commutation relations ${\bf {\bar {\,R}}}\cdot {\bf {\bar S}}={\bf {\bar S}}\cdot\bf {\bar {\,R}}$\cite{dmitriev_book_2002} where ${\bf {\bar {\,R}}}$ are the 2D representations of the generators of the group $C_{4v}$, one comes to the following result:
\begin{equation}
{\bf {\bar S}}=
\left( \!\!
\begin{array} {cc}
 S_{11} &  0 \\[1mm]
 0      &  S_{11}
\end{array} \!\!
\right).
\label{eq:scatC4v}
\end{equation}
The equality $S_{21}=S_{12}=0$ in this matrix indicates that the cross-polarization effect in this array is absent. The result $S_{22}=S_{11}$ defines polarization insensitivity of the metasurface.

The IRREP's of the group $C_{4v}$ are given in Tab.~\ref{tab:C4v}.\cite{hamermesh_book_1962} The upper line of this table shows eight elements of the group. The left column gives five IRREP's according to five classes existing in this group. There are four one-dimensional IRREP's $\Gamma_i, (i=1,...4)$, and one two-dimensional IRREP $\Gamma_5$.

The orthogonal normalized eigenvectors ${\bf E}_1$ and ${\bf E}_2$ of matrix (\ref{eq:scatC4v}) are written as follows:  
\begin{equation}
{\bf E}_1 =
\left(\!
\begin{array}{c}
1 \\
0
\end{array}\!
\right), \qquad
{\bf E}_2 =
\left(\!
\begin{array}{c}
0 \\
1
\end{array}\!
\right),
\label{eq:2.8.17}
\end{equation}
and the corresponding eigenvalues are $s_{1,2}=S_{11}$. These two modes are degenerate ones because they belong to the same IRREP $\Gamma_5$. Thus, due to the symmetry $C_{4v}$, the array  is characterized by polarization degeneracy.

Any linear combination of the two degenerate eigenvectors
${\bf E}_1$ and ${\bf E}_2$, for example, the right and left circular polarized waves
\begin{equation}
{\bf E}_1 \pm i{\bf E}_2
\label{eq:2.8.19}
\end{equation}
are also eigenvectors. These combined vectors are also transformed according to the two-dimensional IRREP $\Gamma_5$ of Tab.~\ref{tab:C4v}.

Tab.~\ref{tab:C4v} comprises transformation properties of all the possible eigenmodes of the array with symmetry $C_{4v}$. However, symmetry of the excited array consisting of the complex system ``individual resonators + square unit cell + excitation field'' includes also symmetry of the external field ${\bf E}$ and its orientation. In particular, the field ${\bf E}$ does not have rotation $C_{4}$ by $\pi/2$ around the $z$-axis. This can reduce the resultant symmetry of the array and the electromagnetic field. For the discussed symmetry $C_{4v}$ and the $x$-polarized, $y$-polarized or $d$-polarized  external field, exclusion of $C_{4}$ reduced the symmetry $C_{4v}$ to the symmetry $C_{2v}$ whose IRREP's  are  given in Tab.~\ref{tab:C2v}. As a result, the field $H_z$ transforms in accordance with the IRREP $\Gamma_4$ of the group $C_{2v}$. This peculiarity is confirmed in numerical simulations of the near-field distribution for the metasurface excited by the $y$-polarized field (see Fig.~\ref{fig:fig4}(b)).

\subsubsection{Numerical results}  
\begin{figure}[htp]
\centering
\includegraphics[width=85mm]{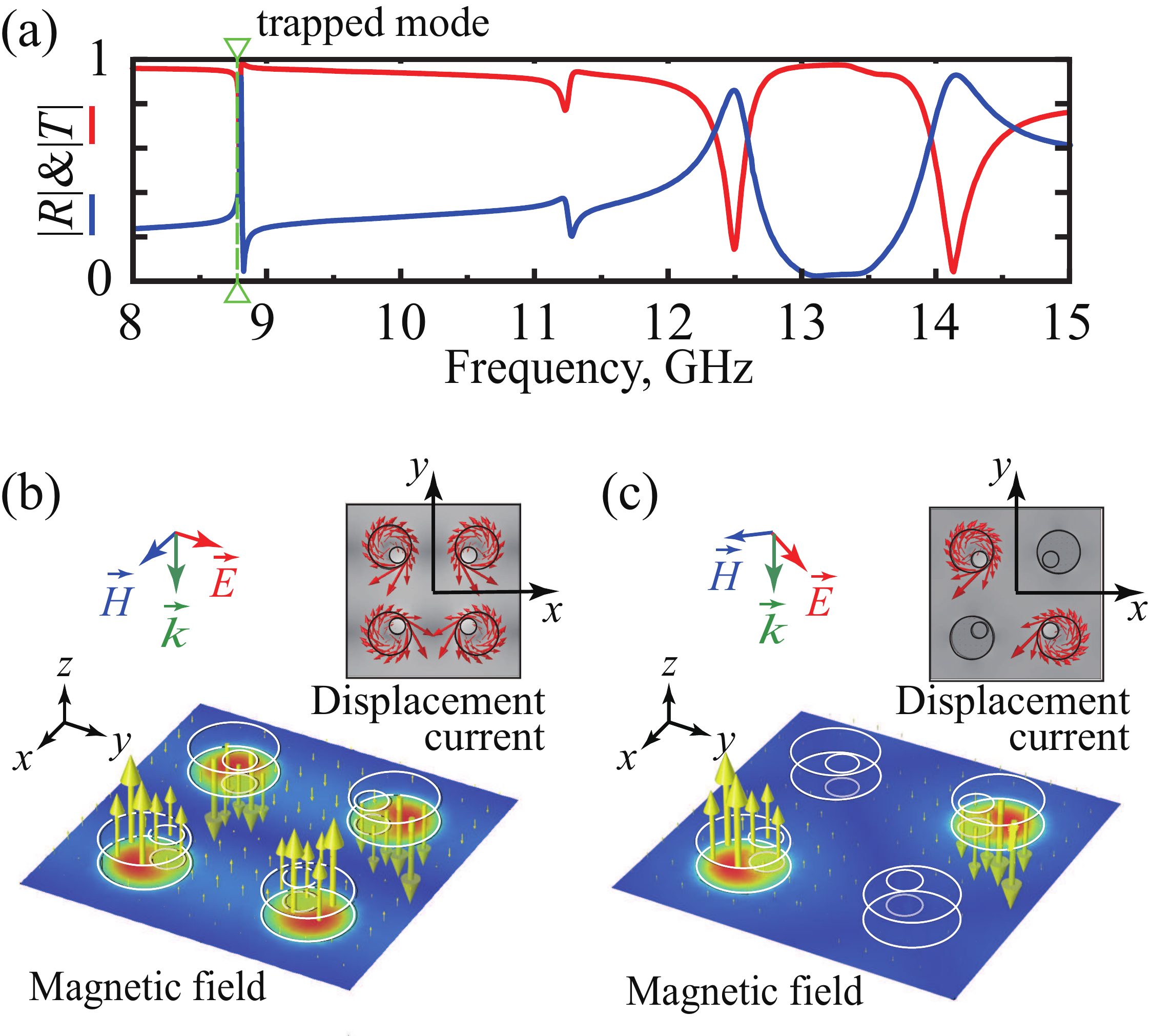}
\caption{(a) Transmission ($T$) and reflection ($R$) coefficients of an all-dielectric metasurface whose unit super-cell is composed of four particles oriented according to the symmetry $C_{4v}$, and distributions of displacement current (red arrows) and magnetic filed (yellow arrows) calculated within the unit super-cell at the resonant frequency of the trapped mode excited by (b) $y$-polarized and (c) $d$-polarized normally incident wave. Geometrical sizes and parameters of the metasurface constituents are: $D_\textrm{disk} = 8.0$ mm, $H_\textrm{disk} = 2.5$ mm, $D_\textrm{hole} = 3.0$ mm, $a = 2.0$ mm, $D = 32$ mm, $H_\textrm{subs} = 20$ mm, $\varepsilon_d = 24$ and $\varepsilon_s = 1.1$.}
\label{fig:fig4}
\end{figure}

In our investigation we relay on both numerical simulation and experimental study. In accordance with our experimental possibilities, we choose the microwave part of the spectrum ($8-15$ GHz) for the metasurface operating. We consider particles made of a microwave ceramic which is characterized by high permittivity and low loss tangent. These particles are equidistantly arranged into a lattice with the period $D$. The lattice is fixed on a dielectric substrate. All geometrical sizes and parameters of the metasurface constituents are summarized in the caption of Fig.~\ref{fig:fig4}.

The numerical simulations of the electromagnetic response of the metasurface are performed with the use of RF module of commercial COMSOL Multiphysics\textsuperscript{\textregistered} finite-element electromagnetic solver. In the solver, the Floquet-periodic boundary conditions are imposed on four sides of the unit cell to simulate the infinite two-dimensional array of resonators.\cite{comsol}
 
The simulated transmitted and reflected spectra of the $C_{4v}$ metasurface are plotted in Fig.~\ref{fig:fig4}(a). Obtained curves are valid for the metasurface irradiation by the wave of an arbitrary linear polarization, since polarization independence is a key feature of this metasurface. The resonant state which corresponds to the trapped mode excitation is marked in the figure by a green arrow. This resonance acquires a sharp peak-and-trough (Fano) profile where extremes of transmission and reflection approach to 0 and 1 alternately, since in the calculations possible intrinsic losses in constitutive materials forming the metasurface are ignored.

Although the spectral curves are the same for all types of linear polarization, the field distribution appears to be different for the metasurface excitation by the waves having $x$- or $y$-polarization and $d$-polarization. The corresponding cross-section patterns (total field) calculated at the resonant frequency of the trapped mode excitation are presented in Figs.~\ref{fig:fig4}(b) and \ref{fig:fig4}(c) for the $y$-polarization and $d$-polarization, respectively. They are plotted in the middle plane at $z$ coordinate corresponding to the half height of the particles. The difference is in the fact, that for the $y$-polarization, all the four resonators within the unit super-cell are active, whereas for the $d$-polarization, only a pair of resonators is active (which resonators are active or inactive depends on along which diagonal of the unit super-cell the electric field vector of excitation is directed). It is revealed that in each active resonator the near-field characteristic resembles that of the TE$_{01\delta}$ mode of the individual cylindrical dielectric resonator discussed above. In both cases the out-of-plane magnetic moments induced in the active particles are oriented in the opposite directions, resembling a dynamic AFM order\cite{Wegener_PhysRevB_2009, Decker_OptLett_2009} (see a detailed discussion on magnetic order in Sec.~\ref{sec:magneticorder}).

\subsection{Array with $C_{2v}$  symmetry}

In order to reach the symmetry $C_{2v}$, orientations of the planes of symmetry $\sigma$ of the perturbed resonators must be along the symmetry elements $\sigma_{x}$ or $\sigma_{y}$ of the unit super-cell (Fig.~\ref{fig:fig3}).

\subsubsection{Scattering matrix}

The calculated reflection matrix for this case is
\begin{equation}
{\bf {\bar S}}=
\left( \!\!
\begin{array} {cc}
S_{11} &  0 \\[1mm]
0 &  S_{22}
\end{array} \!\!
\right),
\label{eq:2.7.7}
\end{equation}
i.e. the reflection coefficients for the $x$- and $y$-polarized waves have different values $(S_{22}\neq S_{11}$), but cross-polarization effect is absent due to equality $(S_{21}=S_{12}=0$). 

%
\begin{table}[!htbp]
	\begin{center}
		\caption{Irreducible representations 
			of the group $C_{2v}$ }
		{
			\begin{tabular}{c@{\hspace{4.0mm}}c@{\hspace{4.0mm}}
					c@{\hspace{4.0mm}}c@{\hspace{4.0mm}}c}
				\hline
				\fbox{$C_{2v}$}
				& e   & $C_2$          & $\sigma_x$     & $\sigma_y$     \\
				\hline
				$\Gamma_1$ & 1   & 1              & 1              & 1               \\
				$\Gamma_2$ & 1   & 1              &$\!\!\!\!\! -1$ &$\!\!\!\!\! -1$ \\
				$\Gamma_3$ & 1   &$\!\!\!\!\! -1$ & 1              &$\!\!\!\!\! -1$  \\
				$\Gamma_4$ & 1   &$\!\!\!\!\! -1$ &$\!\!\!\!\! -1$ & 1       \\
				\\ [-4.0mm] \hline
			\end{tabular}
		}
		\label{tab:C2v}
	\end{center}
\end{table}

The eigenvectors ${\bf E}_1$ and ${\bf E}_2$ can be written
as follows  
\begin{equation}
{\bf E}_1 =
\left(\!
\begin{array}{c}
1 \\
0
\end{array}\!
\right), \qquad
{\bf E}_2 =
\left(\!
\begin{array}{c}
0 \\
1
\end{array}\!
\right).
\label{eq:2.8.17}
\end{equation}
The eigenvalues are $s_{1}=S_{11}$ and $s_{2}=S_{22}$. The eigenvectors ${\bf E}_1$ and ${\bf E}_2$ belong to IRREP's $\Gamma_4$ and $\Gamma_3$ of Tab.~\ref{tab:C2v}, respectively.

Notice, that the group $C_{2v}$ is the direct product of the groups $C_{2}$ and $C_{s}$  discussed below, i.e. $C_{2v}=C_{2}\otimes C_{s}$. Therefore, it combines the properties of the groups $C_{2}$ and $C_{s}$. 
\subsubsection{Numerical results} 
\begin{figure}[htp]
\centering
\includegraphics[width=80mm]{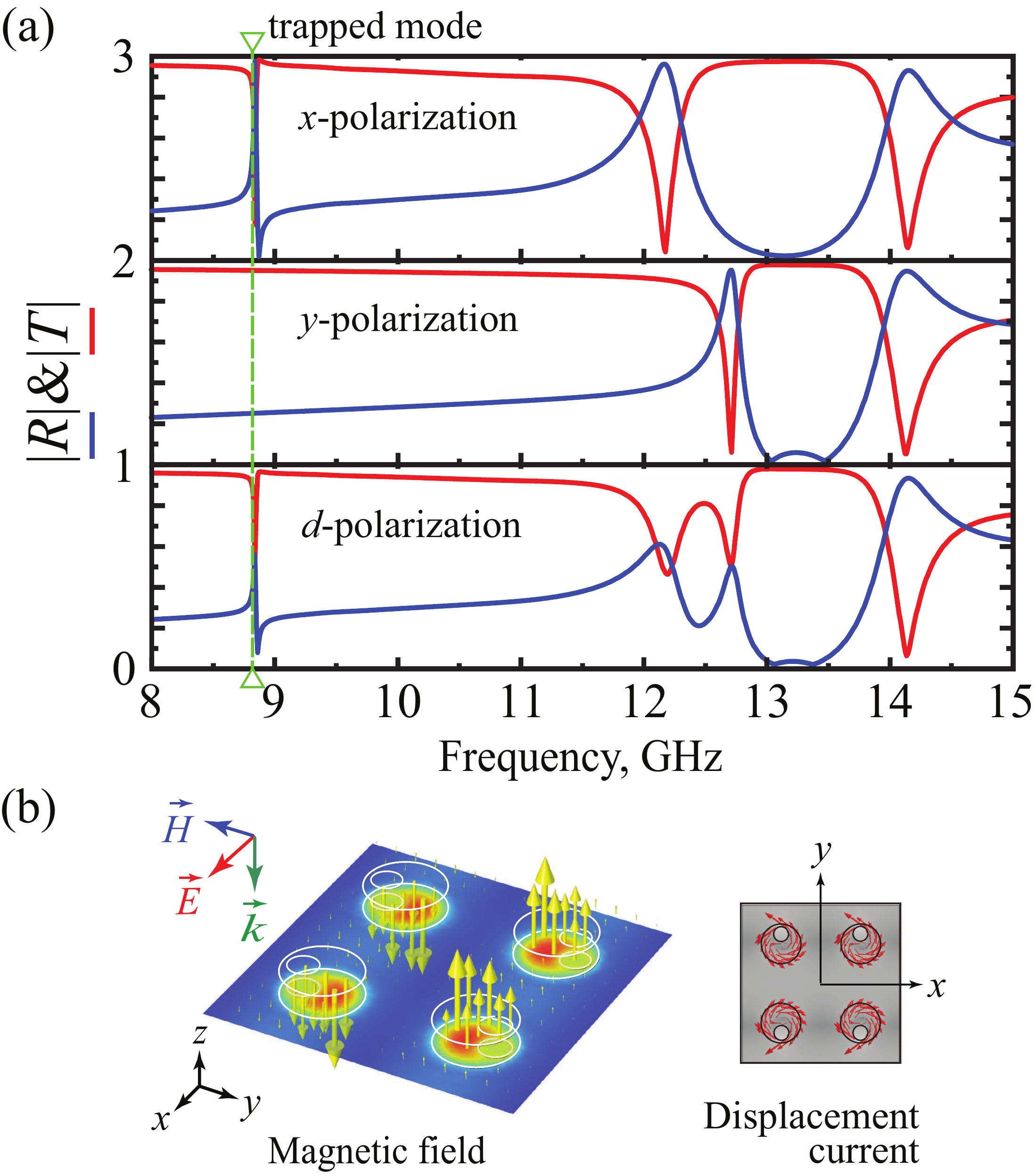}
\caption{Same as in Fig.~\ref{fig:fig4} but for the metasurface whose unit super-cell corresponds to the group $ C_{2v}$ and (b) the metasurface is excited by $x$-polarized wave.}
\label{fig:fig5}
\end{figure}

The simulated transmitted and reflected spectra of the $C_{2v}$ metasurface are presented in Fig.~\ref{fig:fig5}(a). The curves are different for all three polarizations. The trapped mode resonance is absent in the spectra of the metasurface excited by the $y$-polarized wave. For the other two polarizations, the frequency of the trapped mode remains unchanged as compared to the previously discussed case. At the resonant frequency of the trapped mode excitation, the  patterns of the total field are identical for both $x$-polarized and $d$-polarized waves, where four resonators within the unit super-cell are active ones demonstrating AFM order of the induced magnetic moments (Fig.~\ref{fig:fig5}(b)). The field $H_z$ is transformed in accordance with the IRREP  $\Gamma_{4}$ of Tab.~\ref{tab:C2v}.

\subsection{Array with $C_{4}$ symmetry}

For the $C_{4}$ symmetry, two enantiomorphic modifications exist which can be obtained by reflection in a plane of symmetry. These two modifications are denoted in Fig.~\ref{fig:fig3} by symbols $C_4(1)$ and $C_4(2)$.  

\subsubsection{Scattering matrix}

The calculated reflection  matrix for this case is
\begin{equation}
{\bf {\bar S}}=
\left( \,\,
\begin{array} {cc}
S_{11} &  S_{12} \\[1mm]
\!\!\!\!-S_{12}      &  S_{11}
\end{array} \!\!
\right).
\label{eq:2.7.7}
\end{equation}
It shows that the array is polarization independent  due to equality $S_{22} = S_{11}$. Besides, it potentially can exhibit chirality, because $S_{21} = -S_{12}$.

The group  $C_4$ which is a subgroup of $C_{4v}$ contains only rotations around the $z$-axis. It is known from the group theory,\cite{hamermesh_book_1962} that the two-dimensional IRREP $\Gamma_{5}$ of $C_{4v}$ (Tab.~\ref{tab:C4v}) splits into the one-dimensional IRREPs $\Gamma_{3}$ and $\Gamma_{4}$ of the subgroup $C_4$ (Tab.~\ref{tab:C4}).

The  normalized eigenvectors take the following form:
\begin{equation}
{\bf E}_{-1}=\frac{1}{\sqrt{2}}\,
\left(\,
\begin{array}{c}
         1  \\
\!\!\!\!-i  \\
\end{array}  \!
\right), \qquad
{\bf E}_{+1}=\frac{1}{\sqrt{2}}\,
\left(\!
\begin{array}{c}
         1  \\
         i  \\
\end{array} \!
\right),
\label{eq:2.8.26}
\end{equation}
and the eigenvalues are $s_{1,2}=S_{11} \mp S_{12}$. These vectors describe two circularly polarized modes. Notice that they are also eigenvectors of the array with $C_{4v}$ symmetry. But unlike the symmetry $C_{4v}$, the vectors ${\bf E}_{-1}$ and ${\bf E}_{+1}$ of $C_{4}$ are not degenerate ones, because they belong to different one-dimensional IRREP's $\Gamma_{3}$ and $\Gamma_{4}$.

\begin{table}[!htbp]
\begin{center}

	\caption{Irreducible representations of the group $C_4$}	
{
\begin{tabular}{c@{\hspace{4.0mm}}c@{\hspace{4.0mm}}
c@{\hspace{4.0mm}}c@{\hspace{4.0mm}}c}
\hline
\fbox{$C_4$}
           & e   & $C_2$          & $C_4$     & $C_4^{-1}$      \\
\hline     
$\Gamma_1$ & 1   & 1              & 1    &    1           \\
$\Gamma_2$ & 1   & 1              &$\!\!\!\!\! -1$ &$\!\!\!\!\! -1$   \\
$\Gamma_3$ & 1   &$\!\!\!\!\! -1$ & $i$              &$\!\!\!\!\! -i$  \\
$\Gamma_4$ & 1   &$\!\!\!\!\! -1$ &$\!\!\!\!\! -i$ & $i$              \\
\hline
\end{tabular}
}
\label{tab:C4}
\end{center}
\end{table}

\subsubsection{Numerical results}  
\begin{figure}[htp]
\centering
\includegraphics[width=85mm]{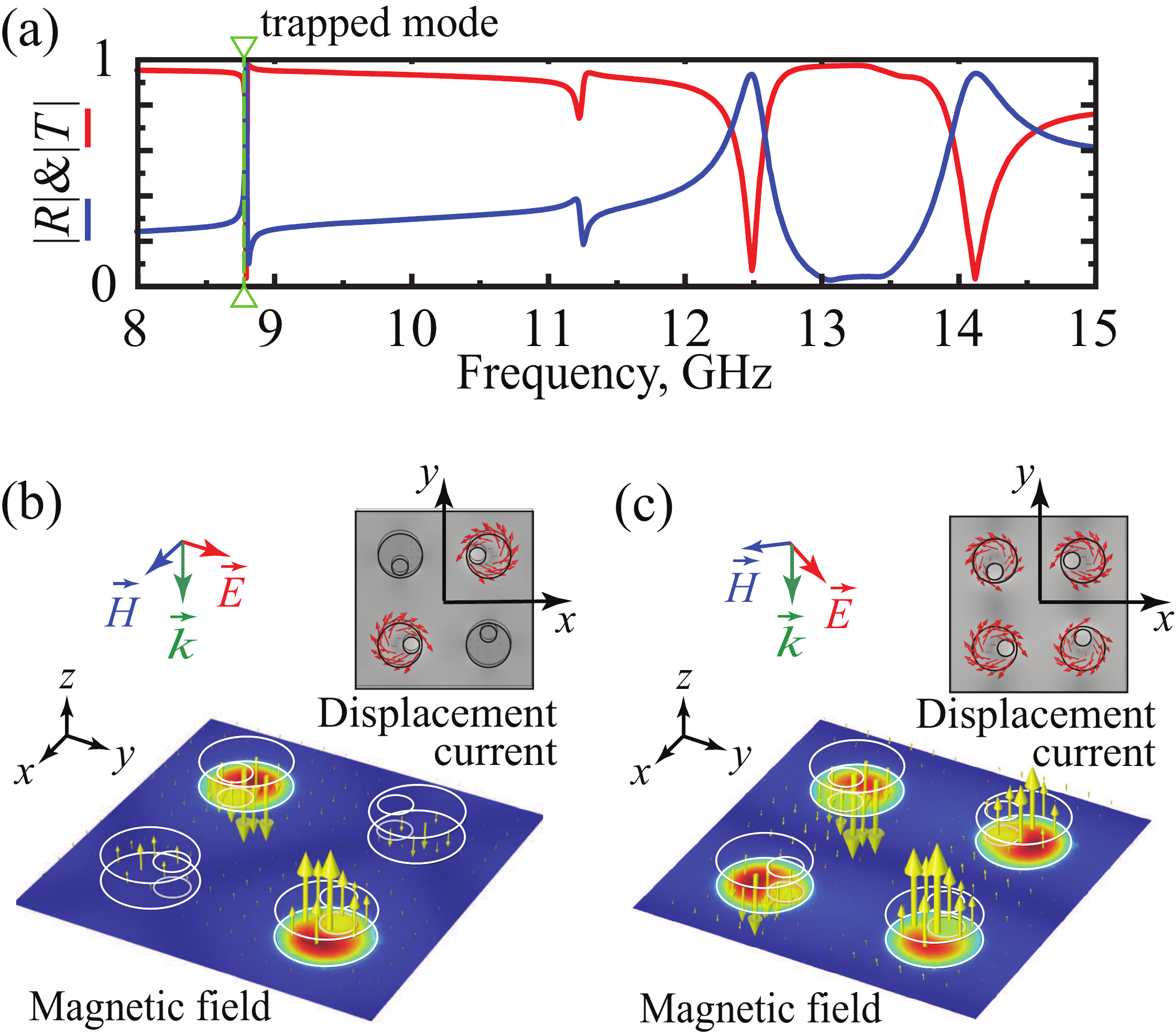}
\caption{Same as in Fig.~\ref{fig:fig4} but for the metasurface whose unit super-cell corresponds to the group $C_{4}$.}
\label{fig:fig6}
\end{figure}

The simulated transmitted and reflected spectra of the $C_{4}$ metasurface are plotted in Fig.~\ref{fig:fig6}(a). One can see that the curves repeat those of the $C_{4v}$ metasurface (Fig.~\ref{fig:fig4}(a)), so the frequency of the trapped mode remains unchanged. The spectral curves are the same for all three cases of the linear polarization of the incident wave. Nevertheless, the difference is in the cross-section patterns of the total field calculated at the frequency of the trapped mode resonance. A pair of active and a pair of inactive resonators occur under the excitation by the $x$- or $y$-polarized wave (Fig.~\ref{fig:fig6}(b)), whereas all four resonators within the unit super-cell are active when the metasurface is excited by the $d$-polarized wave (Fig.~\ref{fig:fig6}(c)).

The same reason as in the case of $C_{4v}$ above, shows that inclusion of symmetry of external excitation reduces the discussed group $C_4$ to $C_2$, such that the field $H_z$ is transformed in accordance to the IRREP $\Gamma_{2}$ of Tab.~\ref{tab:C2}.

\subsection{Array with $C_2$ symmetry}

For the $C_2$ symmetry, two enantiomorphic modifications  also exist which can be obtained by reflection in a plane of symmetry. These two modifications are denoted in Fig.~\ref{fig:fig3} by symbols $C_2(1)$ and $C_2(2)$. 

%
\begin{table}[!htbp]
	\begin{center}
		\caption{Irreducible representations 
			of the group $ C_2$ } 
		{
			\begin{tabular}{c@{\hspace{4.0mm}}c@{\hspace{4.0mm}}c}
				\hline
				\fbox{$C_{2}$}  & $e$   & $C_2$     \\
				\hline
				$\Gamma_1$     & 1   &  1      \\
				$\Gamma_2$     & 1   & $\!\!\!\!-1$   \\
				\hline
			\end{tabular}
		}
		\label{tab:C2}
	\end{center}
\end{table}
%
%
\subsubsection{Scattering matrix}
The calculated   matrix for this case is
\begin{equation}
{\bf {\bar S}}=
\left( \!\!
\begin{array} {cc}
S_{11} &  S_{12} \\[1mm]
S_{21} &  S_{22}
\end{array} \!\!
\right).
\label{eq:2.7.7}
\end{equation}
The scattering matrix  has  a general  form, therefore, the symmetry $C_{2}$  does not give any information about the matrix  and the eigenwave structures. In general, the lower symmetry, the less information one can derive from it. 
\subsubsection{Numerical results} 

\begin{figure}[!htp]
\centering
\includegraphics[width=85mm]{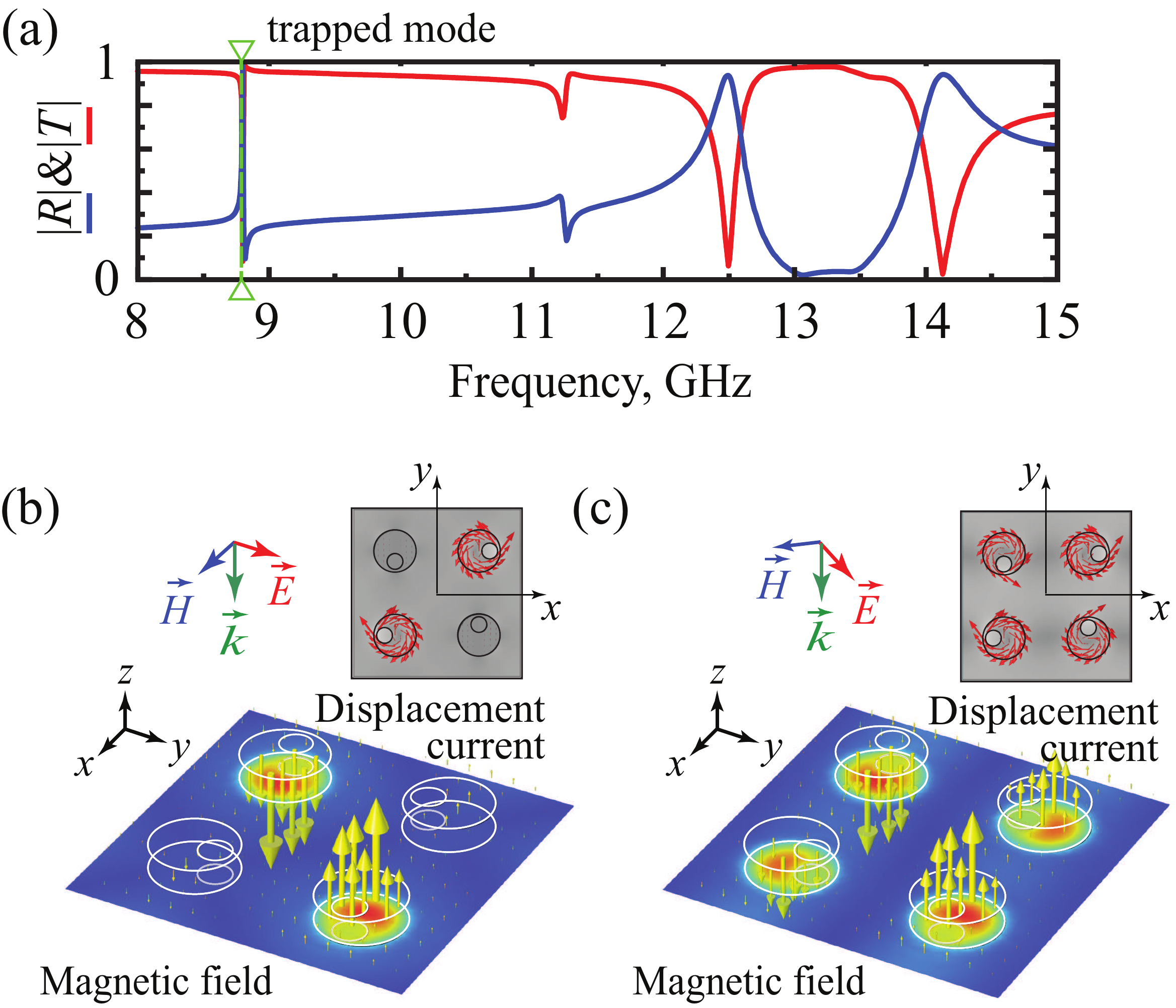}
\caption{Same as in Fig.~\ref{fig:fig4} but for the metasurface whose unit super-cell corresponds to the group $C_2(1)$.}
\label{fig:fig7}
\end{figure}

The simulated transmitted and reflected spectra of the $C_2(1)$ metasurface are plotted in Fig.~\ref{fig:fig7}(a). The curves repeat those of both $C_{4v}$ and $C_{4}$ metasurfaces (see Figs.~\ref{fig:fig4}(a) and \ref{fig:fig6}(a)). Moreover, the cross-section patterns of the total field calculated at the frequency of the trapped mode resonance resemble those of the $C_{4}$ metasurface for the same polarizations of the incident wave (Fig.~\ref{fig:fig7}(b) and \ref{fig:fig7}(c)). The field $H_z$ belongs to the IRREP $\Gamma_{2}$ of Tab.~\ref{tab:C2}.

We do not present here results for the array with symmetry $C_2(2)$, since they are similar to those  of $C_2(1)$. 

\subsection{Array with $C_s$  symmetry}
There are two  modifications for the $C_{s}$ group which contain only one plane of symmetry $\sigma_y$ or $\sigma_x$. They are denoted in Fig.~\ref{fig:fig3} by symbols $C_s(1)$ and $C_s(2)$, respectively. For the $C_s(1)$ group all disk are equally oriented forming the unit super-cell symmetric with respect to  the plane $y=0$. Such geometry in the  metasurface has been considered earlier in several papers.\cite{jain_advoptmater_2015, Tuz_OptExpress_2018, Cui_acsphotonics_2018}  Another modification corresponding to the $C_s(2)$ group has the orientation of particles, where in a pair of disks, the holes are oriented inward of the unit super-cell, whereas in an another pair of disks, the holes are oriented outward. For this case, the unit super-cell is  symmetric with respect to  the plane  $x=0$.
\subsubsection{Scattering matrix}
The calculated reflection  matrix for this case is
\begin{equation}
{\bf {\bar S}}=
\left( \!\!
\begin{array} {cc}
S_{11} &  0 \\[1mm]
0 &  S_{22}
\end{array} \!\!
\right),
\label{eq:2.7.7}
\end{equation}
i.e. the reflection coefficients for $x$- and $y$-polarized incident waves have different values $(S_{22}\neq S_{11}$), but cross-polarization effect is absent $(S_{21}=S_{12}=0$). Notice that the structure of the scattering matrix for ${C}_{s}$ coincides with that of $C_{2v}$ symmetry.
%

\begin{table}[!htbp]
	\begin{center}
		\caption{Irreducible representations 
			of the group $ C_s$ } 
		{
			\begin{tabular}{c@{\hspace{4.0mm}}c@{\hspace{4.0mm}}c}
				\hline
				\fbox{$C_{s}$}  & $e$   & $\sigma_{C_s}$      \\
				\hline
				$\Gamma_1$     & 1   &  1            \\
				$\Gamma_2$     & 1   & $\!\!\!\!-1$   \\
				\hline
			\end{tabular}
		}
		\label{tab:Cs}
	\end{center}
\end{table}

%
The eigenvectors ${\bf E}_1$ and ${\bf E}_2$ are
\begin{equation}
{\bf E}_1 =
\left(\!
\begin{array}{c}
1 \\
0
\end{array}\!
\right), \qquad
{\bf E}_2 =
\left(\!
\begin{array}{c}
0 \\
1
\end{array}\!
\right),
\label{eq:2.8.17}
\end{equation}
and the corresponding eigenvalues are $s_{1}=S_{11}$ and $s_{2}=S_{22}$.
Belonging of ${\bf E}_1$ and ${\bf E}_2$
to one or another IRREP depends on polarization  of the incident wave   with respect to the plane $\sigma_{C_s}$.
\subsubsection{Numerical results} 
\begin{figure}[htp]
\centering
\includegraphics[width=85mm]{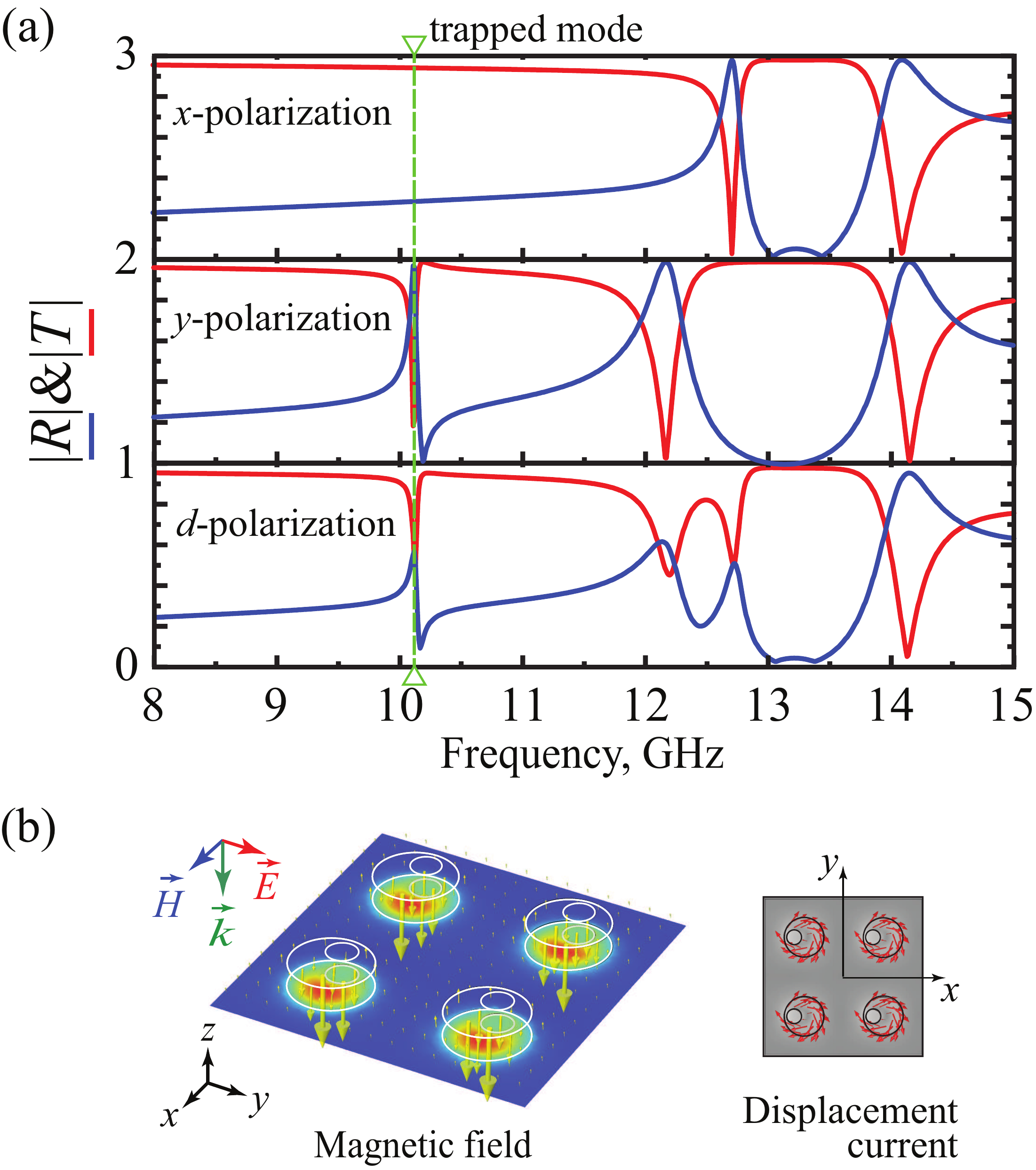}
\caption{Same as in Fig.~\ref{fig:fig4} but for the metasurface whose unit super-cell corresponds to the group $ C_s(1)$.}
\label{fig:fig8}
\end{figure}

In the $C_s(1)$ metasurface, for the chosen orientation of the disks within the unit super-cell, the trapped mode appears to be excited only by the $y$- and $d$-polarized waves (Fig.~\ref{fig:fig8}(a)). Curves of the transmitted and reflected spectra are different for all three polarizations. The resonant frequency of the trapped mode appears to by shifted to the higher frequency band compared to the previously discussed cases. 

At the frequency of the trapped mode excitation, there is a specific distribution of the electromagnetic field within each particle, where the out-of-plane magnetic moments are oriented in the same direction in all resonators demonstrating the FM order (Fig.~\ref{fig:fig8}(b)). The field $H_z$ belongs to the IRREP $\Gamma_{2}$ of Tab.~\ref{tab:Cs}.

\begin{figure}[htp]
\centering
\includegraphics[width=85mm]{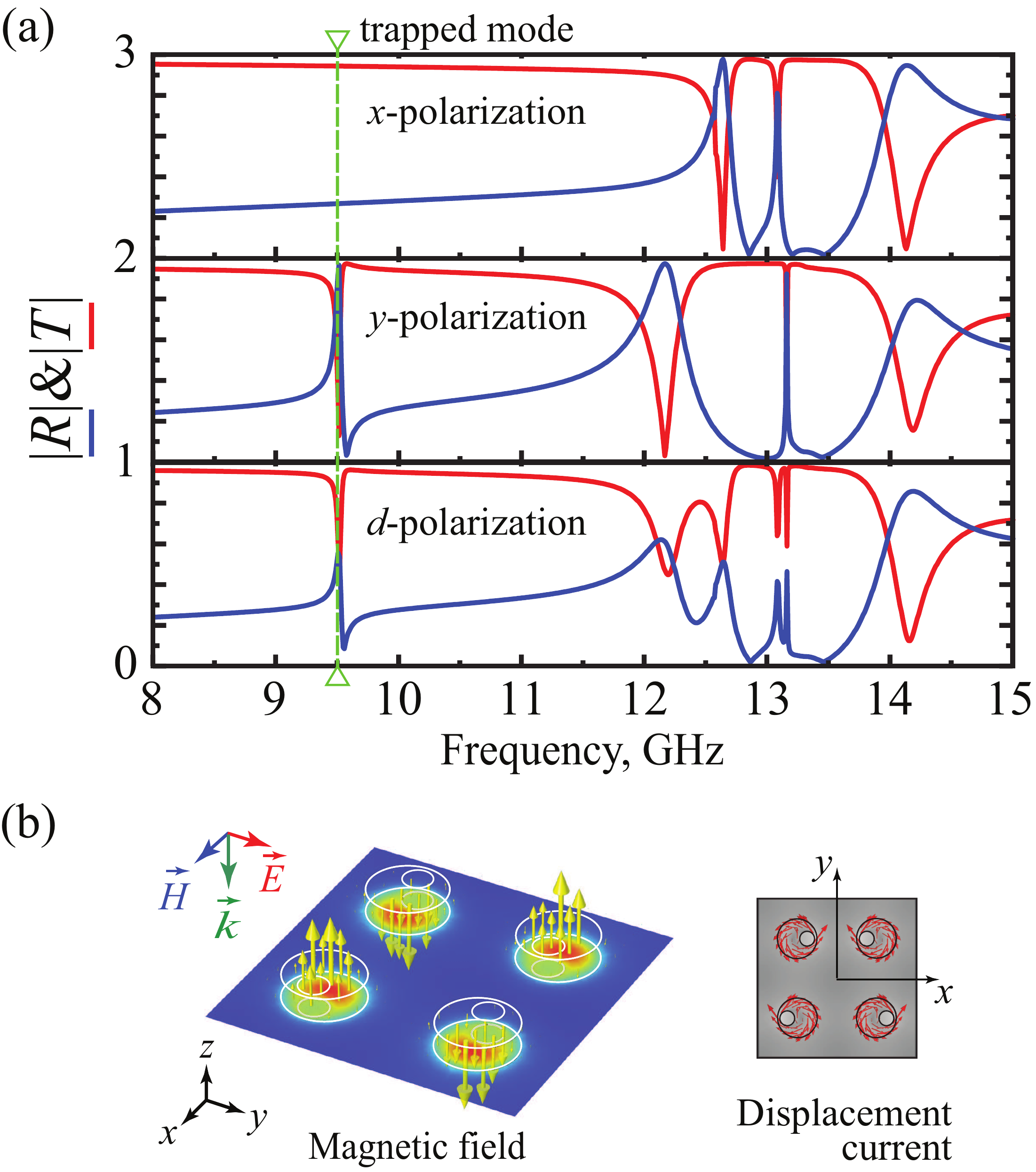}
\caption{Same as in Fig.~\ref{fig:fig4} but for the metasurface whose unit super-cell corresponds to the group $ C_s(2)$.}
\label{fig:fig9}
\end{figure}

The results for the $C_s(2)$ metasurface are completely different from those of the $C_s(1)$ metasurface, where again curves of the transmitted and reflected spectra are different for all three polarizations (Fig.~\ref{fig:fig9}(a)). The trapped mode appears to be excited only by the $y$- and $d$-polarized waves. The resonant frequency of the trapped mode coincides with all previously discussed cases, except the $C_s(1)$ metasurface. At the frequency of the trapped mode excitation, the out-of-plane magnetic moments are oriented up and down in the pairs of particles disposed in the unit super-cell diagonals demonstrating the AFM order (Fig.~\ref{fig:fig9}(b)). In this case, the field $H_z$ belongs to the IRREP  $\Gamma_1$ of Tab.~\ref{tab:Cs}.
\section{Symmetry conditions for AFM and FM orders}
\label{sec:magneticorder}
\subsection{Theoretical description}

As shown above, in the metasurface under study two phase states appear: (i) the dynamic FM order, when all out-of-plane magnetic moments induced in resonators by the incident wave are oriented in the same direction, and (ii) the dynamic AFM order, where the magnetic moments in one half of resonators within the unit super-cell are oriented up, whereas those in the another half of resonators are oriented down. Here a general question arises: how symmetries are related to the corresponding magnetic order? 

It is revealed that the resonators orientation within the unit super-cell corresponding to the symmetries $C_{4v}$, $C_{2v}$, $C_{4}$, and $C_{2}$ result in the dynamic AFM order, whereas only the arrays possessing the symmetry $C_{s}$ can demonstrate both FM and AFM orders. The  transition from the FM order to the AFM order and vice versa can be fulfilled by simple rotation of the individual resonators around their axes.

The phases of the field in the resonators induced by the external excitation depends on the orientation of holes in the resonators forming the unit super-cell. Existence of the rotation by $\pi$ in a group corresponds to such orientation of the holes in the corresponding symmetrical resonators, that the excitation of the related by this symmetry disks produces the displacement currents in them in opposite directions. This gives the anti-parallel orientation of the magnetic dipole fields. Therefore, such a symmetry of the unit super-cell produces the AFM order. The groups $C_{4v}$, $C_{4}$, $C_{2v}$, and $C_{2}$ contain  rotation by $\pi$. Thus, excitation of the FM order in arrays with these symmetries is impossible. This is similar to selection rules known from the quantum mechanics.\cite{hamermesh_book_1962} 

The unique group which does not contain rotation by $\pi$ is the group $C_{s}$ with only one plane of symmetry. Two variants of geometry for this group were  considered above. One of them is $C_s(1)$ with orientation of the plane $\sigma$  parallel to the plane $\sigma_x$. It leads to the FM order of array (Fig.~\ref{fig:fig8}). Another one is $C_s(2)$ corresponding to  orientation of the plane $\sigma$ perpendicular to $\sigma_y$. This gives the AFM order (Fig.~\ref{fig:fig9}). Thus, one can conclude, that for the unit super-cell of the metasurface  described by the group $C_{s}$, both FM and AFM orders of the array are allowed. 

The change in symmetry of a crystal plays an essential role in the Landau's theory of the second-order phase transitions.\cite{landau_1960_5} Analogously to this theory, in our case one can consider transition of the state with high symmetry $C_{4v}$ for the array with non-perturbed resonators (``nonmagnetic state'') to the state with lower symmetries where the array with perturbed resonators behaving either FM or AFM order. The continuous parameter of the Landau's theory describing changing in transition can be, for example, the diameter of the hole or its position with respect to the center of the resonator, or the angle of disk rotation about its axis.

\subsection{Experimental verification}

In order to verify the appearance of magnetic orders at the trapped mode excitation, several samples of the metasurface were fabricated for their characterization in the frequency range of $8-15$~GHz. For the experimental study, the $C_s(1)$ and $C_s(2)$ metasurfaces are chosen since they can demonstrate different magnetic order. We used cylindrical particles made from a commercially available Taizhou Wangling TP-series ceramic (dielectric loss tangent is $\tan\delta \approx 0.015$ at $10$~GHz). The substrate was made of a Styrofoam. The set of particles was prepared by using the mechanical cutting technique. The overall size of the sample is approximately $600\times600$~mm, and it consists of an array of $20\times20$ particles. The rest parameters of the metasurface correspond to those given in the caption of Fig.~\ref{fig:fig4}.

\begin{figure*}[htp]
\centering
\includegraphics[width=170mm]{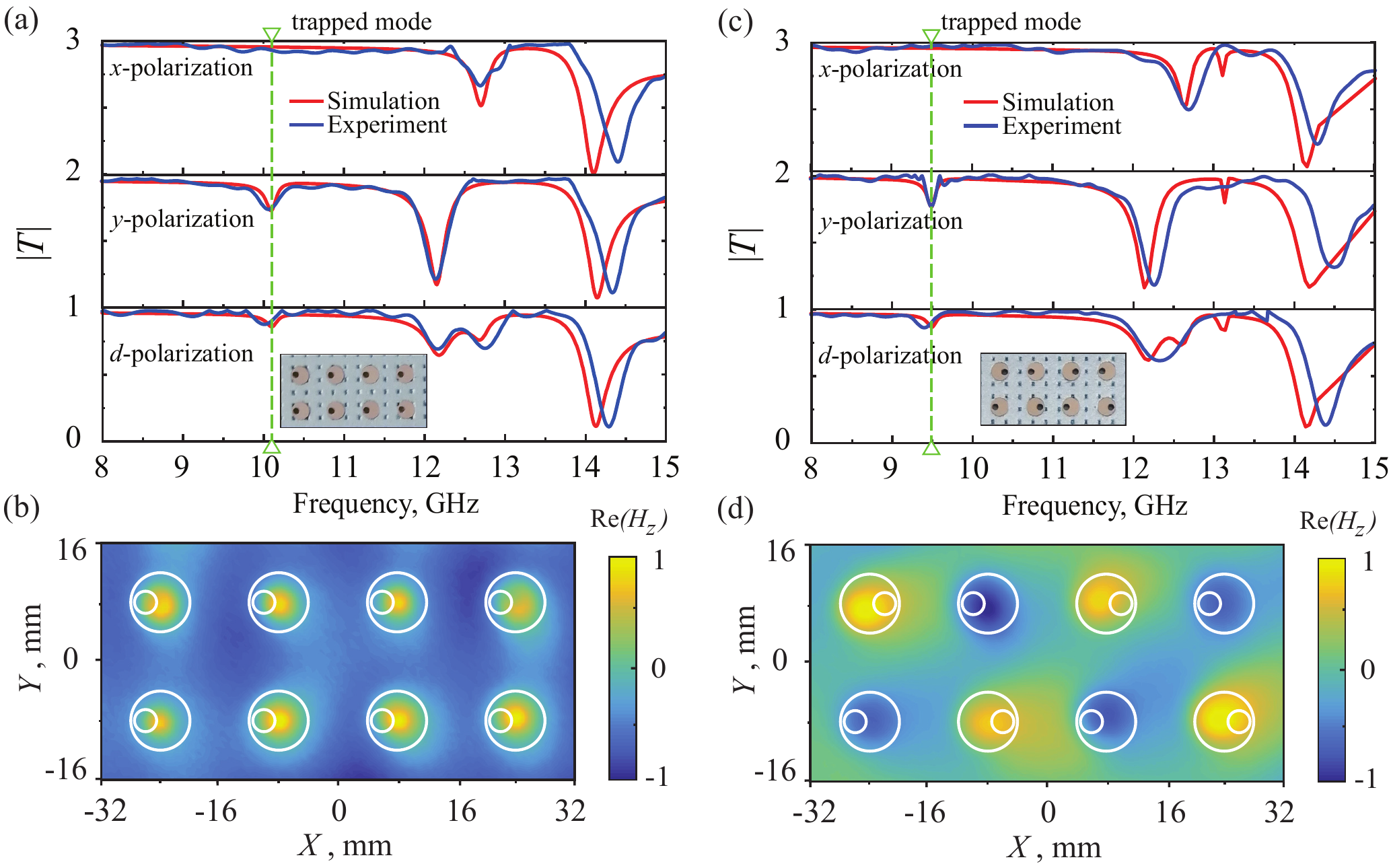}
\caption{(a, c) Simulated (red lines) and measured (blue lines) transmission coefficient, and (b, d) measured real part of the $H_z$ component of the magnetic near-field (out-of-plane magnetic moments) at the corresponding resonant frequency of trapped mode excited by the $y$-polarized incident wave. The insets demonstrate fragments of the metasurface prototypes. In the simulation actual material losses ($\tan\delta=0.015$) in ceramic disks are taken into account, while the substrate is modeled as a lossless dielectric. All other parameters of the metasurface are the same as in Fig.~\ref{fig:fig4}. (a, b) $C_s(1)$ and (c, d) $C_s(2)$.}
\label{fig:fig10}
\end{figure*}

Using Vector Network Analyzer, the $S_{21}$-parameter (transmission coefficient) of the samples was measured for three different polarizations ($x$-, $y$-, and $d$-polarization) of the incident wave. The spectral characteristics of the metasurface were measured in an anechoic chamber. We used a custom made post-processing procedure to remove unwanted noise from the measured data. The measured and simulated results are given in Figs.~\ref{fig:fig10}(a) and \ref{fig:fig10}(b) for the $C_s(1)$ and $C_s(2)$ metasurfaces, respectively. In the numerical simulation we have accounted for actual losses inherent to the ceramic used for the disks (for details on the measurement setup and samples preparation, see Refs.~\onlinecite{tuz_AdvOptMat_2019} and \onlinecite{sayanskiy_arXiv_2018}). 

The magnetic orders can be experimentally studied from the measurement of the near-field distribution of the metasurfaces at the frequency of the trapped mode excitation. According to the results of our numerical simulations at this frequency the electric field is mostly concentrated inside the particles, whereas the magnetic field is penetrating out of the resonators. Moreover, the magnetic moments are orthogonally oriented to the metasurface plane and demonstrate different patterns with respect to the particular resonator orientation. Thus, we have measured the magnetic near-field distribution for the proposed designs. The color maps of the measured real part of the $H_z$ component of the near-field confirming discussed above trapped mode resonant conditions and orientations of the out-of-plane magnetic moments are shown in Figs.~\ref{fig:fig10}(c) and \ref{fig:fig10}(d) for the $C_s(1)$ and $C_s(2)$ metasurfaces, respectively.

\section{Interaction of resonators}
\label{sec:coupling}
The coupling of the TE$_{01\delta}$ mode  of the cylindrical resonator with the field of irradiating wave is provided by the hole made in the resonator. It is a principal mechanism of excitation. Besides, interaction of resonators occurs due to coupling of the neighbours by evanescent fields excited around every resonator by the incident wave. The proximity of the resonators provides coherent interaction between them. 

The  interaction of the resonators due to their near fields is defined by the distance between them and by the permittivity of the resonator material. In our case, the existence of such interaction is confirmed, in particular, by difference in the resonant frequencies for the FM and AFM orders (compare Figs.~\ref{fig:fig8}(a) and \ref{fig:fig9}(a)). The FM order leads to reduction of the electric fields between the resonators because of their opposite directions. It results in higher concentration of the fields inside the resonators. As a consequence, the resonant frequency shifts to the higher frequency band.  Contrariwise, the AFM order increases field between the resonators, which reduces the resonant frequency.

Mathematically speaking, the interaction of the neighboring resonator is defined by integral of superposition of their fields. The force of interaction can not be defined by the symmetry arguments. But in our case, the symmetry can help to evaluate, at least, qualitatively, this interaction. If both of the two mechanisms of excitation are present (i.e. due to the external field and due to the mutual coupling), the level of excitation of the resonators depend on the phases of these two excitation, i.e. on their constructive or destructive interaction in the resonator. 

The coupling between resonators, in general, can also exist due to the electric dipole-dipole, and higher multipole interactions which, in general, produce much lower effects. The influence of the higher modes depends greatly on the proximity of their resonances to the resonance of the principal TE$_{01\delta}$ mode.

Analysis of the mechanism of excitation due to mutual coupling of the resonators requires a more systematic study. This problem is out of scope of this paper, and can be a subject for a future work.

\section{Polarization dependence of resonator excitation}
\label{sec:polarization}
It was shown above, that the arrays with the symmetries $C_{4v}$ and  $C_{4}$ are polarization insensitive, but the excitation of the individual particles depends on the polarization of the incident field. For example, one can see in Fig.~\ref{fig:fig4}, that in the case of symmetry $C_{4v}$ and $x$- or $y$-polarization, all the four particles within the unit super-cell are excited, but for the $d$-polarization, only two of them are active. In the case of symmetry $C_{4}$, the situation is inverted. 

However, the total power stored in the resonators is almost equal in all these four cases. To show this, let us consider the array with symmetry $C_{4v}$ presented in Fig.~\ref{fig:fig4}. For a given excitation power, the electric field intensity perpendicular to the  plane of symmetry $\sigma$ in the hole of the resonator for the case of $x$-polarization is lower than that for the case of $d$-polarization (compare Figs.~\ref{fig:fig4}(b) and \ref{fig:fig4}(c)). The $x$-polarized field $\bf E$   can be decomposed into the components parallel and perpendicular to the plane $\sigma$. For the excitation with the amplitude $E_x$, the field intensity perpendicular to the plane $\sigma$ is $E_x/\sqrt{2}$, i.e. the excitation power of each of the four resonators is two times lower than that of the excitation of any of the two resonators for the diagonal excitation. Thus, for the resonators with sufficiently small holes, the overall intensity of the array excitation is approximately the same for these two polarization of the incident field. In the same manner, one can analyze arrays with other symmetries.

\section{Conclusions}
\label{sec:concl}

The obtained above results can be considered as a convenient classification scheme for the symmetrical arrays with four resonators in a unit cell. This scheme is based on the possible symmetries and comprises the highest group $C_{4v}$ and four its subgroups. 

The eigenwaves and the fields of every array belong to one or another IRREP of the corresponding group and they have definite transformation properties. The presented symmetry analysis, in particular, the exact scattering matrices, can be helpful in numerical calculus to reduce the computation time and the necessary volume of memory, and also in experiments, for example, to validate the obtained results.

The effects related to magnetization waves and their dynamics, such as retardation effects, 
\cite{Wegener_PhysRevB_2009, Decker_OptLett_2009}
analysis of Ising model,\cite{Singh_AdvMat_2018} etc. discussed for arrays with metallic split ring resonators can be studied using dielectric arrays presented above. These effects can be more pronounced due to lower losses in dielectric resonators in comparison to those inherent to metallic resonators at higher frequencies.

The suggested method can be applied to resonators with other types of dark modes, as well as to arrays with other symmetries,  such as, for example, $C_{6}$, $C_{6v}$ with six resonators in a unit super-cell. 

\section*{\label{ack}Acknowledgments}
VD thanks  Brazilian agency CNPq for  financial support, VRT acknowledges Jilin University's hospitality and financial support.


\section*{Appendix A: Scattering cross section of a perturbed particle}
\label{sec:appendix}
\renewcommand{\theequation}{A\arabic{equation}}
\setcounter{equation}{0}
\renewcommand{\thefigure}{A\arabic{figure}}
\setcounter{figure}{0}   

We performed additional simulations to study a scattering characteristic of an isolated perturbed resonator (see Fig.~\ref{fig:fig2}). As an exciting radiation, a plane wave having linear polarization is supposed. We consider incidence of waves of both orthogonal polarizations ($x$-polarized and $y$-polarized waves). The wave incidents along the axis of the cylinder (frontal excitation). The sum of the contributions from different multipole moments is written as\cite{Jackson_1998}
\begin{equation}
\begin{split}
C_\textrm{sca}^\textrm{total} = C_\textrm{sca}^p + C_\textrm{sca}^m + C_\textrm{sca}^{Q_e} + C_\textrm{sca}^{Q_m} +...,
\end{split}
\label{eq:expansion}
\end{equation}
where, $C_\textrm{sca}^p$, $C_\textrm{sca}^m$, $C_\textrm{sca}^{Q_e}$, $C_\textrm{sca}^{Q_m}$ are the contributions to scattering cross section from electric dipole (ED), magnetic dipole (MD), electric quadrupole (EQ) and magnetic quadrupole, respectively.

\begin{figure}[htp]
\centering
\includegraphics[width=67mm]{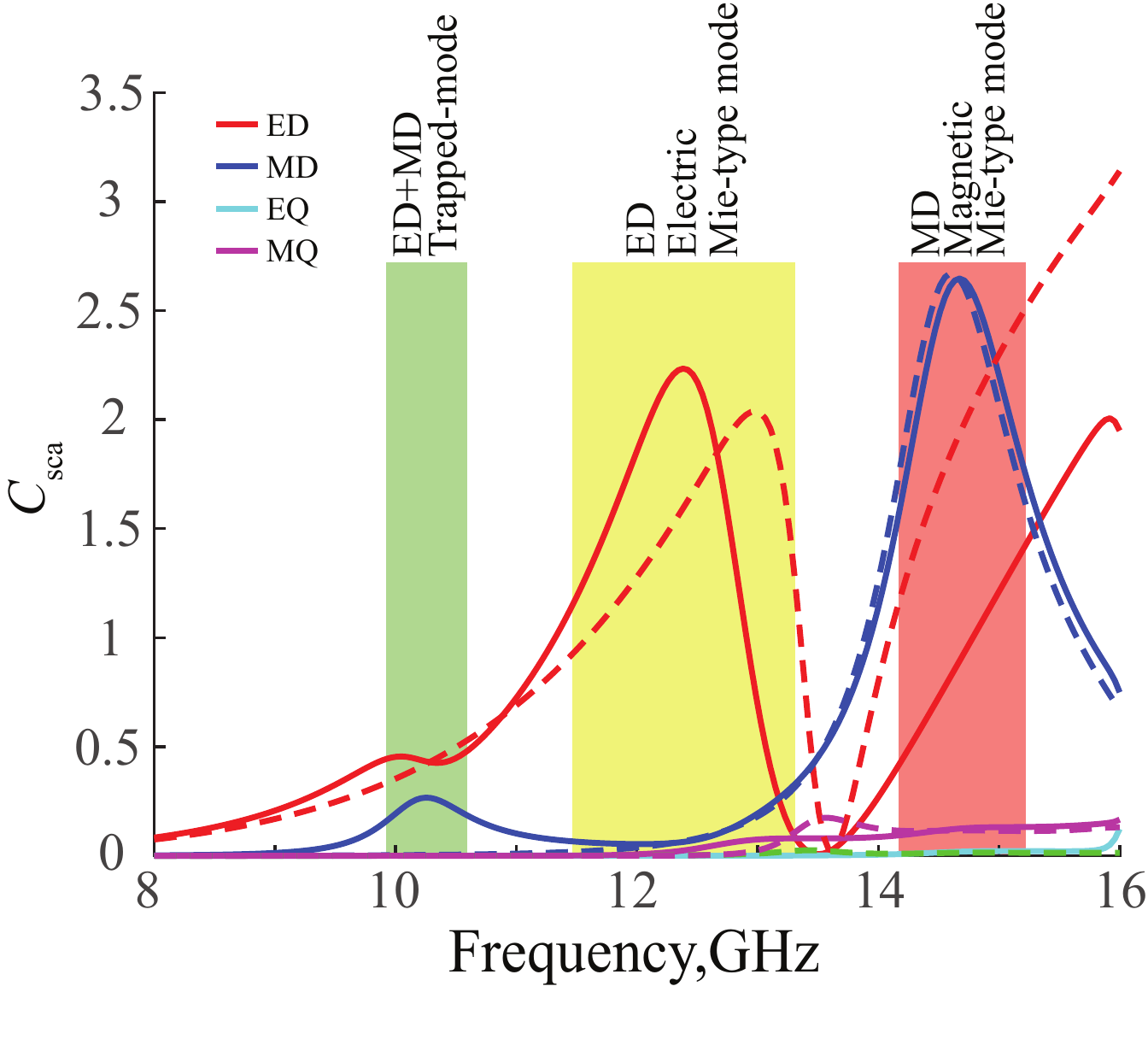}
\caption{Normalized (normalized by the value $\lambda^2/2\pi$) contributions of four lowest order multipole moments to the scattering cross section of a single particle irradiated by either $x$-polarized (solid lines) or $y$-polarized (dashed lines) wave. The areas colored in green, yellow and red correspond to frequency bands where the trapped mode, ED Mie-type mode, and MD Mie-type mode exist, respectively.}
\label{fig:appendix}
\end{figure}

In Fig.~\ref{fig:appendix} the contribution of four lowest-order multipole (dipole and quadrupole) moments to the scattering cross section of the given particle irradiated by the $x$-polarized and $y$-polarized wave are shown. Along with resonances related to the lowest-order ED and MD Mie-type modes,\cite{Baryshnikova_JOSA_B_2017} an additional resonant state appears in the scattering cross section of the particle (see green area in Fig.~\ref{fig:appendix}). It only appears in the response on excitation by the $x$-polarized wave, so it is peculiar to the perturbed particle having an eccentric hole. It is related to the excitation of the trapped mode since it is a feature of asymmetric systems.\cite{Fedotov_PhysRevLett_2007} One can see that both ED and MD moments contribute to this resonance. 
\bigskip 
\bibliography{trapped_modes}

\begin{thebibliography}{49}%
\makeatletter
\providecommand \@ifxundefined [1]{%
 \@ifx{#1\undefined}
}%
\providecommand \@ifnum [1]{%
 \ifnum #1\expandafter \@firstoftwo
 \else \expandafter \@secondoftwo
 \fi
}%
\providecommand \@ifx [1]{%
 \ifx #1\expandafter \@firstoftwo
 \else \expandafter \@secondoftwo
 \fi
}%
\providecommand \natexlab [1]{#1}%
\providecommand \enquote  [1]{``#1''}%
\providecommand \bibnamefont  [1]{#1}%
\providecommand \bibfnamefont [1]{#1}%
\providecommand \citenamefont [1]{#1}%
\providecommand \href@noop [0]{\@secondoftwo}%
\providecommand \href [0]{\begingroup \@sanitize@url \@href}%
\providecommand \@href[1]{\@@startlink{#1}\@@href}%
\providecommand \@@href[1]{\endgroup#1\@@endlink}%
\providecommand \@sanitize@url [0]{\catcode `\\12\catcode `\$12\catcode
  `\&12\catcode `\#12\catcode `\^12\catcode `\_12\catcode `\%12\relax}%
\providecommand \@@startlink[1]{}%
\providecommand \@@endlink[0]{}%
\providecommand \url  [0]{\begingroup\@sanitize@url \@url }%
\providecommand \@url [1]{\endgroup\@href {#1}{\urlprefix }}%
\providecommand \urlprefix  [0]{URL }%
\providecommand \Eprint [0]{\href }%
\providecommand \doibase [0]{http://dx.doi.org/}%
\providecommand \selectlanguage [0]{\@gobble}%
\providecommand \bibinfo  [0]{\@secondoftwo}%
\providecommand \bibfield  [0]{\@secondoftwo}%
\providecommand \translation [1]{[#1]}%
\providecommand \BibitemOpen [0]{}%
\providecommand \bibitemStop [0]{}%
\providecommand \bibitemNoStop [0]{.\EOS\space}%
\providecommand \EOS [0]{\spacefactor3000\relax}%
\providecommand \BibitemShut  [1]{\csname bibitem#1\endcsname}%
\let\auto@bib@innerbib\@empty
\bibitem [{\citenamefont {Kruk}\ and\ \citenamefont
  {Kivshar}(2017)}]{kruk_acsphotonics_2017}%
  \BibitemOpen
  \bibfield  {author} {\bibinfo {author} {\bibfnamefont {S.}~\bibnamefont
  {Kruk}}\ and\ \bibinfo {author} {\bibfnamefont {Y.}~\bibnamefont {Kivshar}},\
  }\href {\doibase 10.1021/acsphotonics.7b01038} {\bibfield  {journal}
  {\bibinfo  {journal} {ACS Photonics}\ }\textbf {\bibinfo {volume} {4}},\
  \bibinfo {pages} {2638} (\bibinfo {year} {2017})}\BibitemShut {NoStop}%
\bibitem [{\citenamefont {Zhao}\ \emph {et~al.}(2009)\citenamefont {Zhao},
  \citenamefont {Zhou}, \citenamefont {Zhang},\ and\ \citenamefont
  {Lippens}}]{Zhao_MaterToday_2009}%
  \BibitemOpen
  \bibfield  {author} {\bibinfo {author} {\bibfnamefont {Q.}~\bibnamefont
  {Zhao}}, \bibinfo {author} {\bibfnamefont {J.}~\bibnamefont {Zhou}}, \bibinfo
  {author} {\bibfnamefont {F.}~\bibnamefont {Zhang}}, \ and\ \bibinfo {author}
  {\bibfnamefont {D.}~\bibnamefont {Lippens}},\ }\href {\doibase
  10.1016/S1369-7021(09)70318-9} {\bibfield  {journal} {\bibinfo  {journal}
  {Mater. Today}\ }\textbf {\bibinfo {volume} {12}},\ \bibinfo {pages} {60}
  (\bibinfo {year} {2009})}\BibitemShut {NoStop}%
\bibitem [{\citenamefont {Jahani}\ and\ \citenamefont
  {Jacob}(2016)}]{Jahani_Nanotechnol_2016}%
  \BibitemOpen
  \bibfield  {author} {\bibinfo {author} {\bibfnamefont {S.}~\bibnamefont
  {Jahani}}\ and\ \bibinfo {author} {\bibfnamefont {Z.}~\bibnamefont {Jacob}},\
  }\href {\doibase 10.1038/nnano.2015.304} {\bibfield  {journal} {\bibinfo
  {journal} {Nat. Nanotechnol}\ }\textbf {\bibinfo {volume} {11}},\ \bibinfo
  {pages} {23} (\bibinfo {year} {2016})}\BibitemShut {NoStop}%
\bibitem [{\citenamefont {Limonov}\ \emph {et~al.}(2017)\citenamefont
  {Limonov}, \citenamefont {Rybin}, \citenamefont {Poddubny},\ and\
  \citenamefont {Kivshar}}]{Limonov_NatPhoton_2017}%
  \BibitemOpen
  \bibfield  {author} {\bibinfo {author} {\bibfnamefont {M.~F.}\ \bibnamefont
  {Limonov}}, \bibinfo {author} {\bibfnamefont {M.~V.}\ \bibnamefont {Rybin}},
  \bibinfo {author} {\bibfnamefont {A.~N.}\ \bibnamefont {Poddubny}}, \ and\
  \bibinfo {author} {\bibfnamefont {Y.~S.}\ \bibnamefont {Kivshar}},\ }\href
  {\doibase 10.1038/nphoton.2017.142} {\bibfield  {journal} {\bibinfo
  {journal} {Nat. Photonics}\ }\textbf {\bibinfo {volume} {11}},\ \bibinfo
  {pages} {543} (\bibinfo {year} {2017})}\BibitemShut {NoStop}%
\bibitem [{\citenamefont {Baranov}\ \emph {et~al.}(2017)\citenamefont
  {Baranov}, \citenamefont {Zuev}, \citenamefont {Lepeshov}, \citenamefont
  {Kotov}, \citenamefont {Krasnok}, \citenamefont {Evlyukhin},\ and\
  \citenamefont {Chichkov}}]{Baranov_Optica_2017}%
  \BibitemOpen
  \bibfield  {author} {\bibinfo {author} {\bibfnamefont {D.~G.}\ \bibnamefont
  {Baranov}}, \bibinfo {author} {\bibfnamefont {D.~A.}\ \bibnamefont {Zuev}},
  \bibinfo {author} {\bibfnamefont {S.~I.}\ \bibnamefont {Lepeshov}}, \bibinfo
  {author} {\bibfnamefont {O.~V.}\ \bibnamefont {Kotov}}, \bibinfo {author}
  {\bibfnamefont {A.~E.}\ \bibnamefont {Krasnok}}, \bibinfo {author}
  {\bibfnamefont {A.~B.}\ \bibnamefont {Evlyukhin}}, \ and\ \bibinfo {author}
  {\bibfnamefont {B.~N.}\ \bibnamefont {Chichkov}},\ }\href {\doibase
  10.1364/OPTICA.4.000814} {\bibfield  {journal} {\bibinfo  {journal} {Optica}\
  }\textbf {\bibinfo {volume} {4}},\ \bibinfo {pages} {814} (\bibinfo {year}
  {2017})}\BibitemShut {NoStop}%
\bibitem [{\citenamefont {Yang}\ \emph {et~al.}(2015)\citenamefont {Yang},
  \citenamefont {Wang}, \citenamefont {Boulesbaa}, \citenamefont {Kravchenko},
  \citenamefont {Briggs}, \citenamefont {Puretzky}, \citenamefont {Geohegan},\
  and\ \citenamefont {Valentine}}]{Yang_NanoLett_2015}%
  \BibitemOpen
  \bibfield  {author} {\bibinfo {author} {\bibfnamefont {Y.}~\bibnamefont
  {Yang}}, \bibinfo {author} {\bibfnamefont {W.}~\bibnamefont {Wang}}, \bibinfo
  {author} {\bibfnamefont {A.}~\bibnamefont {Boulesbaa}}, \bibinfo {author}
  {\bibfnamefont {I.~I.}\ \bibnamefont {Kravchenko}}, \bibinfo {author}
  {\bibfnamefont {D.~P.}\ \bibnamefont {Briggs}}, \bibinfo {author}
  {\bibfnamefont {A.}~\bibnamefont {Puretzky}}, \bibinfo {author}
  {\bibfnamefont {D.}~\bibnamefont {Geohegan}}, \ and\ \bibinfo {author}
  {\bibfnamefont {J.}~\bibnamefont {Valentine}},\ }\href {\doibase
  10.1021/acs.nanolett.5b02802} {\bibfield  {journal} {\bibinfo  {journal}
  {Nano Lett.}\ }\textbf {\bibinfo {volume} {15}},\ \bibinfo {pages} {7388}
  (\bibinfo {year} {2015})}\BibitemShut {NoStop}%
\bibitem [{\citenamefont {Shcherbakov}\ \emph
  {et~al.}(2015{\natexlab{a}})\citenamefont {Shcherbakov}, \citenamefont
  {Shorokhov}, \citenamefont {Neshev}, \citenamefont {Hopkins}, \citenamefont
  {Staude}, \citenamefont {Melik-Gaykazyan}, \citenamefont {Ezhov},
  \citenamefont {Miroshnichenko}, \citenamefont {Brener}, \citenamefont
  {Fedyanin},\ and\ \citenamefont {Kivshar}}]{Shcherbakov_acsphotonics_2015}%
  \BibitemOpen
  \bibfield  {author} {\bibinfo {author} {\bibfnamefont {M.~R.}\ \bibnamefont
  {Shcherbakov}}, \bibinfo {author} {\bibfnamefont {A.~S.}\ \bibnamefont
  {Shorokhov}}, \bibinfo {author} {\bibfnamefont {D.~N.}\ \bibnamefont
  {Neshev}}, \bibinfo {author} {\bibfnamefont {B.}~\bibnamefont {Hopkins}},
  \bibinfo {author} {\bibfnamefont {I.}~\bibnamefont {Staude}}, \bibinfo
  {author} {\bibfnamefont {E.~V.}\ \bibnamefont {Melik-Gaykazyan}}, \bibinfo
  {author} {\bibfnamefont {A.~A.}\ \bibnamefont {Ezhov}}, \bibinfo {author}
  {\bibfnamefont {A.~E.}\ \bibnamefont {Miroshnichenko}}, \bibinfo {author}
  {\bibfnamefont {I.}~\bibnamefont {Brener}}, \bibinfo {author} {\bibfnamefont
  {A.~A.}\ \bibnamefont {Fedyanin}}, \ and\ \bibinfo {author} {\bibfnamefont
  {Y.~S.}\ \bibnamefont {Kivshar}},\ }\href {\doibase
  10.1021/acsphotonics.5b00065} {\bibfield  {journal} {\bibinfo  {journal} {ACS
  Photonics}\ }\textbf {\bibinfo {volume} {2}},\ \bibinfo {pages} {578}
  (\bibinfo {year} {2015}{\natexlab{a}})}\BibitemShut {NoStop}%
\bibitem [{\citenamefont {Liu}\ \emph {et~al.}(2016)\citenamefont {Liu},
  \citenamefont {Sinclair}, \citenamefont {Saravi}, \citenamefont {Keeler},
  \citenamefont {Yang}, \citenamefont {Reno}, \citenamefont {Peake},
  \citenamefont {Setzpfandt}, \citenamefont {Staude}, \citenamefont {Pertsch},\
  and\ \citenamefont {Brener}}]{Liu_NanoLett_2016}%
  \BibitemOpen
  \bibfield  {author} {\bibinfo {author} {\bibfnamefont {S.}~\bibnamefont
  {Liu}}, \bibinfo {author} {\bibfnamefont {M.~B.}\ \bibnamefont {Sinclair}},
  \bibinfo {author} {\bibfnamefont {S.}~\bibnamefont {Saravi}}, \bibinfo
  {author} {\bibfnamefont {G.~A.}\ \bibnamefont {Keeler}}, \bibinfo {author}
  {\bibfnamefont {Y.}~\bibnamefont {Yang}}, \bibinfo {author} {\bibfnamefont
  {J.}~\bibnamefont {Reno}}, \bibinfo {author} {\bibfnamefont {G.~M.}\
  \bibnamefont {Peake}}, \bibinfo {author} {\bibfnamefont {F.}~\bibnamefont
  {Setzpfandt}}, \bibinfo {author} {\bibfnamefont {I.}~\bibnamefont {Staude}},
  \bibinfo {author} {\bibfnamefont {T.}~\bibnamefont {Pertsch}}, \ and\
  \bibinfo {author} {\bibfnamefont {I.}~\bibnamefont {Brener}},\ }\href
  {\doibase 10.1021/acs.nanolett.6b01816} {\bibfield  {journal} {\bibinfo
  {journal} {Nano Lett.}\ }\textbf {\bibinfo {volume} {16}},\ \bibinfo {pages}
  {5426} (\bibinfo {year} {2016})}\BibitemShut {NoStop}%
\bibitem [{\citenamefont {Shcherbakov}\ \emph
  {et~al.}(2015{\natexlab{b}})\citenamefont {Shcherbakov}, \citenamefont
  {Vabishchevich}, \citenamefont {Shorokhov}, \citenamefont {Chong},
  \citenamefont {Choi}, \citenamefont {Staude}, \citenamefont {Miroshnichenko},
  \citenamefont {Neshev}, \citenamefont {Fedyanin},\ and\ \citenamefont
  {Kivshar}}]{Shcherbakov_Nanolett_2015}%
  \BibitemOpen
  \bibfield  {author} {\bibinfo {author} {\bibfnamefont {M.~R.}\ \bibnamefont
  {Shcherbakov}}, \bibinfo {author} {\bibfnamefont {P.~P.}\ \bibnamefont
  {Vabishchevich}}, \bibinfo {author} {\bibfnamefont {A.~S.}\ \bibnamefont
  {Shorokhov}}, \bibinfo {author} {\bibfnamefont {K.~E.}\ \bibnamefont
  {Chong}}, \bibinfo {author} {\bibfnamefont {D.-Y.}\ \bibnamefont {Choi}},
  \bibinfo {author} {\bibfnamefont {I.}~\bibnamefont {Staude}}, \bibinfo
  {author} {\bibfnamefont {A.~E.}\ \bibnamefont {Miroshnichenko}}, \bibinfo
  {author} {\bibfnamefont {D.~N.}\ \bibnamefont {Neshev}}, \bibinfo {author}
  {\bibfnamefont {A.~A.}\ \bibnamefont {Fedyanin}}, \ and\ \bibinfo {author}
  {\bibfnamefont {Y.~S.}\ \bibnamefont {Kivshar}},\ }\href {\doibase
  10.1021/acs.nanolett.5b02989} {\bibfield  {journal} {\bibinfo  {journal}
  {Nano Lett.}\ }\textbf {\bibinfo {volume} {15}},\ \bibinfo {pages} {6985}
  (\bibinfo {year} {2015}{\natexlab{b}})}\BibitemShut {NoStop}%
\bibitem [{\citenamefont {Fu}\ \emph {et~al.}(2013)\citenamefont {Fu},
  \citenamefont {Kuznetsov}, \citenamefont {Miroshnichenko}, \citenamefont
  {Yu},\ and\ \citenamefont {Luk'yanchuk}}]{Fu_NatCommun_2013}%
  \BibitemOpen
  \bibfield  {author} {\bibinfo {author} {\bibfnamefont {Y.~H.}\ \bibnamefont
  {Fu}}, \bibinfo {author} {\bibfnamefont {A.~I.}\ \bibnamefont {Kuznetsov}},
  \bibinfo {author} {\bibfnamefont {A.~E.}\ \bibnamefont {Miroshnichenko}},
  \bibinfo {author} {\bibfnamefont {Y.~F.}\ \bibnamefont {Yu}}, \ and\ \bibinfo
  {author} {\bibfnamefont {B.}~\bibnamefont {Luk'yanchuk}},\ }\href {\doibase
  10.1038/ncomms2538} {\bibfield  {journal} {\bibinfo  {journal} {Nat.
  Commun.}\ }\textbf {\bibinfo {volume} {4}},\ \bibinfo {pages} {1527}
  (\bibinfo {year} {2013})}\BibitemShut {NoStop}%
\bibitem [{\citenamefont {Ginn}\ \emph {et~al.}(2012)\citenamefont {Ginn},
  \citenamefont {Brener}, \citenamefont {Peters}, \citenamefont {Wendt},
  \citenamefont {Stevens}, \citenamefont {Hines}, \citenamefont {Basilio},
  \citenamefont {Warne}, \citenamefont {Ihlefeld}, \citenamefont {Clem},\ and\
  \citenamefont {Sinclair}}]{Ginn_PhysRevLett_2012}%
  \BibitemOpen
  \bibfield  {author} {\bibinfo {author} {\bibfnamefont {J.~C.}\ \bibnamefont
  {Ginn}}, \bibinfo {author} {\bibfnamefont {I.}~\bibnamefont {Brener}},
  \bibinfo {author} {\bibfnamefont {D.~W.}\ \bibnamefont {Peters}}, \bibinfo
  {author} {\bibfnamefont {J.~R.}\ \bibnamefont {Wendt}}, \bibinfo {author}
  {\bibfnamefont {J.~O.}\ \bibnamefont {Stevens}}, \bibinfo {author}
  {\bibfnamefont {P.~F.}\ \bibnamefont {Hines}}, \bibinfo {author}
  {\bibfnamefont {L.~I.}\ \bibnamefont {Basilio}}, \bibinfo {author}
  {\bibfnamefont {L.~K.}\ \bibnamefont {Warne}}, \bibinfo {author}
  {\bibfnamefont {J.~F.}\ \bibnamefont {Ihlefeld}}, \bibinfo {author}
  {\bibfnamefont {P.~G.}\ \bibnamefont {Clem}}, \ and\ \bibinfo {author}
  {\bibfnamefont {M.~B.}\ \bibnamefont {Sinclair}},\ }\href {\doibase
  10.1103/PhysRevLett.108.097402} {\bibfield  {journal} {\bibinfo  {journal}
  {Phys. Rev. Lett.}\ }\textbf {\bibinfo {volume} {108}},\ \bibinfo {pages}
  {097402} (\bibinfo {year} {2012})}\BibitemShut {NoStop}%
\bibitem [{\citenamefont {Kuznetsov}\ \emph {et~al.}(2012)\citenamefont
  {Kuznetsov}, \citenamefont {Miroshnichenko}, \citenamefont {Fu},
  \citenamefont {Zhang},\ and\ \citenamefont
  {Luk'yanchuk}}]{Kuznetsov_SciRep_2012}%
  \BibitemOpen
  \bibfield  {author} {\bibinfo {author} {\bibfnamefont {A.~I.}\ \bibnamefont
  {Kuznetsov}}, \bibinfo {author} {\bibfnamefont {A.~E.}\ \bibnamefont
  {Miroshnichenko}}, \bibinfo {author} {\bibfnamefont {Y.~H.}\ \bibnamefont
  {Fu}}, \bibinfo {author} {\bibfnamefont {J.}~\bibnamefont {Zhang}}, \ and\
  \bibinfo {author} {\bibfnamefont {B.}~\bibnamefont {Luk'yanchuk}},\ }\href
  {\doibase 10.1038/srep00492} {\bibfield  {journal} {\bibinfo  {journal} {Sci.
  Rep.}\ }\textbf {\bibinfo {volume} {2}},\ \bibinfo {pages} {492} (\bibinfo
  {year} {2012})}\BibitemShut {NoStop}%
\bibitem [{\citenamefont {He}\ \emph {et~al.}(2018)\citenamefont {He},
  \citenamefont {Guo}, \citenamefont {Feng}, \citenamefont {Xu},\ and\
  \citenamefont {Miroshnichenko}}]{Miroshnichenko_PhysRevB_2018}%
  \BibitemOpen
  \bibfield  {author} {\bibinfo {author} {\bibfnamefont {Y.}~\bibnamefont
  {He}}, \bibinfo {author} {\bibfnamefont {G.}~\bibnamefont {Guo}}, \bibinfo
  {author} {\bibfnamefont {T.}~\bibnamefont {Feng}}, \bibinfo {author}
  {\bibfnamefont {Y.}~\bibnamefont {Xu}}, \ and\ \bibinfo {author}
  {\bibfnamefont {A.~E.}\ \bibnamefont {Miroshnichenko}},\ }\href {\doibase
  10.1103/PhysRevB.98.161112} {\bibfield  {journal} {\bibinfo  {journal} {Phys.
  Rev. B}\ }\textbf {\bibinfo {volume} {98}},\ \bibinfo {pages} {161112}
  (\bibinfo {year} {2018})}\BibitemShut {NoStop}%
\bibitem [{\citenamefont {Koshelev}\ \emph {et~al.}(2018)\citenamefont
  {Koshelev}, \citenamefont {Lepeshov}, \citenamefont {Liu}, \citenamefont
  {Bogdanov},\ and\ \citenamefont {Kivshar}}]{Koshelev_PhysRevLett_2018}%
  \BibitemOpen
  \bibfield  {author} {\bibinfo {author} {\bibfnamefont {K.}~\bibnamefont
  {Koshelev}}, \bibinfo {author} {\bibfnamefont {S.}~\bibnamefont {Lepeshov}},
  \bibinfo {author} {\bibfnamefont {M.}~\bibnamefont {Liu}}, \bibinfo {author}
  {\bibfnamefont {A.}~\bibnamefont {Bogdanov}}, \ and\ \bibinfo {author}
  {\bibfnamefont {Y.}~\bibnamefont {Kivshar}},\ }\href {\doibase
  10.1103/PhysRevLett.121.193903} {\bibfield  {journal} {\bibinfo  {journal}
  {Phys. Rev. Lett.}\ }\textbf {\bibinfo {volume} {121}},\ \bibinfo {pages}
  {193903} (\bibinfo {year} {2018})}\BibitemShut {NoStop}%
\bibitem [{\citenamefont {Miroshnichenko}\ \emph {et~al.}(2015)\citenamefont
  {Miroshnichenko}, \citenamefont {Evlyukhin}, \citenamefont {Bakker},
  \citenamefont {Chipouline}, \citenamefont {Kuznetsov}, \citenamefont
  {Luk'yanchuk}, \citenamefont {Chichkov},\ and\ \citenamefont
  {Kivshar}}]{Miroshnichenko_NatCommun_2015}%
  \BibitemOpen
  \bibfield  {author} {\bibinfo {author} {\bibfnamefont {A.~E.}\ \bibnamefont
  {Miroshnichenko}}, \bibinfo {author} {\bibfnamefont {A.~B.}\ \bibnamefont
  {Evlyukhin}}, \bibinfo {author} {\bibfnamefont {R.~M.}\ \bibnamefont
  {Bakker}}, \bibinfo {author} {\bibfnamefont {A.}~\bibnamefont {Chipouline}},
  \bibinfo {author} {\bibfnamefont {A.~I.}\ \bibnamefont {Kuznetsov}}, \bibinfo
  {author} {\bibfnamefont {B.}~\bibnamefont {Luk'yanchuk}}, \bibinfo {author}
  {\bibfnamefont {B.~N.}\ \bibnamefont {Chichkov}}, \ and\ \bibinfo {author}
  {\bibfnamefont {Y.~S.}\ \bibnamefont {Kivshar}},\ }\href {\doibase
  10.1038/ncomms9069} {\bibfield  {journal} {\bibinfo  {journal} {Nat.
  Commun.}\ }\textbf {\bibinfo {volume} {6}},\ \bibinfo {pages} {8069}
  (\bibinfo {year} {2015})}\BibitemShut {NoStop}%
\bibitem [{\citenamefont {Luk'yanchuk}\ \emph {et~al.}(2017)\citenamefont
  {Luk'yanchuk}, \citenamefont {Paniagua-Dom\'{\i}nguez}, \citenamefont
  {Kuznetsov}, \citenamefont {Miroshnichenko},\ and\ \citenamefont
  {Kivshar}}]{Lukyanchuk_PhysRevA_2017}%
  \BibitemOpen
  \bibfield  {author} {\bibinfo {author} {\bibfnamefont {B.}~\bibnamefont
  {Luk'yanchuk}}, \bibinfo {author} {\bibfnamefont {R.}~\bibnamefont
  {Paniagua-Dom\'{\i}nguez}}, \bibinfo {author} {\bibfnamefont {A.~I.}\
  \bibnamefont {Kuznetsov}}, \bibinfo {author} {\bibfnamefont {A.~E.}\
  \bibnamefont {Miroshnichenko}}, \ and\ \bibinfo {author} {\bibfnamefont
  {Y.~S.}\ \bibnamefont {Kivshar}},\ }\href {\doibase
  10.1103/PhysRevA.95.063820} {\bibfield  {journal} {\bibinfo  {journal} {Phys.
  Rev. A}\ }\textbf {\bibinfo {volume} {95}},\ \bibinfo {pages} {063820}
  (\bibinfo {year} {2017})}\BibitemShut {NoStop}%
\bibitem [{\citenamefont {Sayanskiy}\ \emph
  {et~al.}(2018{\natexlab{a}})\citenamefont {Sayanskiy}, \citenamefont
  {Danaeifar}, \citenamefont {Kapitanova},\ and\ \citenamefont
  {Miroshnichenko}}]{Kapitanova_AdvOptMat_2018}%
  \BibitemOpen
  \bibfield  {author} {\bibinfo {author} {\bibfnamefont {A.}~\bibnamefont
  {Sayanskiy}}, \bibinfo {author} {\bibfnamefont {M.}~\bibnamefont
  {Danaeifar}}, \bibinfo {author} {\bibfnamefont {P.}~\bibnamefont
  {Kapitanova}}, \ and\ \bibinfo {author} {\bibfnamefont {A.~E.}\ \bibnamefont
  {Miroshnichenko}},\ }\href {\doibase 10.1002/adom.201800302} {\bibfield
  {journal} {\bibinfo  {journal} {Adv. Opt. Mater.}\ }\textbf {\bibinfo
  {volume} {6}},\ \bibinfo {pages} {1800302} (\bibinfo {year}
  {2018}{\natexlab{a}})}\BibitemShut {NoStop}%
\bibitem [{\citenamefont {Tuz}, \citenamefont {Khardikov},\ and\ \citenamefont
  {Kivshar}(2018)}]{tuz_ACSPhotonics_2018}%
  \BibitemOpen
  \bibfield  {author} {\bibinfo {author} {\bibfnamefont {V.~R.}\ \bibnamefont
  {Tuz}}, \bibinfo {author} {\bibfnamefont {V.~V.}\ \bibnamefont {Khardikov}},
  \ and\ \bibinfo {author} {\bibfnamefont {Y.~S.}\ \bibnamefont {Kivshar}},\
  }\href {\doibase 10.1021/acsphotonics.8b00098} {\bibfield  {journal}
  {\bibinfo  {journal} {ACS Photonics}\ }\textbf {\bibinfo {volume} {5}},\
  \bibinfo {pages} {1871} (\bibinfo {year} {2018})}\BibitemShut {NoStop}%
\bibitem [{\citenamefont {Xu}\ \emph {et~al.}(2018)\citenamefont {Xu},
  \citenamefont {Sayanskiy}, \citenamefont {Kupriianov}, \citenamefont {Tuz},
  \citenamefont {Kapitanova}, \citenamefont {Sun}, \citenamefont {Han},\ and\
  \citenamefont {Kivshar}}]{tuz_AdvOptMat_2019}%
  \BibitemOpen
  \bibfield  {author} {\bibinfo {author} {\bibfnamefont {S.}~\bibnamefont
  {Xu}}, \bibinfo {author} {\bibfnamefont {A.}~\bibnamefont {Sayanskiy}},
  \bibinfo {author} {\bibfnamefont {A.~S.}\ \bibnamefont {Kupriianov}},
  \bibinfo {author} {\bibfnamefont {V.~R.}\ \bibnamefont {Tuz}}, \bibinfo
  {author} {\bibfnamefont {P.}~\bibnamefont {Kapitanova}}, \bibinfo {author}
  {\bibfnamefont {H.-B.}\ \bibnamefont {Sun}}, \bibinfo {author} {\bibfnamefont
  {W.}~\bibnamefont {Han}}, \ and\ \bibinfo {author} {\bibfnamefont {Y.~S.}\
  \bibnamefont {Kivshar}},\ }\href {\doibase 10.1002/adom.201801166} {\bibfield
   {journal} {\bibinfo  {journal} {Adv. Opt. Mater.}\ }\textbf {\bibinfo
  {volume} {0}},\ \bibinfo {pages} {1801166} (\bibinfo {year}
  {2018})}\BibitemShut {NoStop}%
\bibitem [{\citenamefont {Lu}, \citenamefont {Joannopoulos},\ and\
  \citenamefont {Solja\v{c}i\'{c}}(2014)}]{Lu_NatPhotonics_2014}%
  \BibitemOpen
  \bibfield  {author} {\bibinfo {author} {\bibfnamefont {L.}~\bibnamefont
  {Lu}}, \bibinfo {author} {\bibfnamefont {J.~D.}\ \bibnamefont
  {Joannopoulos}}, \ and\ \bibinfo {author} {\bibfnamefont {M.}~\bibnamefont
  {Solja\v{c}i\'{c}}},\ }\href {\doibase 10.1038/nphoton.2014.248} {\bibfield
  {journal} {\bibinfo  {journal} {Nat. Photonics}\ }\textbf {\bibinfo {volume}
  {8}},\ \bibinfo {pages} {821} (\bibinfo {year} {2014})}\BibitemShut {NoStop}%
\bibitem [{\citenamefont {Slobozhanyuk}\ \emph {et~al.}(2017)\citenamefont
  {Slobozhanyuk}, \citenamefont {Mousavi}, \citenamefont {Ni}, \citenamefont
  {Smirnova}, \citenamefont {Kivshar},\ and\ \citenamefont
  {Khanikaev}}]{Slobozhanyuk_NatPhotonics_2017}%
  \BibitemOpen
  \bibfield  {author} {\bibinfo {author} {\bibfnamefont {A.}~\bibnamefont
  {Slobozhanyuk}}, \bibinfo {author} {\bibfnamefont {S.~H.}\ \bibnamefont
  {Mousavi}}, \bibinfo {author} {\bibfnamefont {X.}~\bibnamefont {Ni}},
  \bibinfo {author} {\bibfnamefont {D.}~\bibnamefont {Smirnova}}, \bibinfo
  {author} {\bibfnamefont {Y.~S.}\ \bibnamefont {Kivshar}}, \ and\ \bibinfo
  {author} {\bibfnamefont {A.~B.}\ \bibnamefont {Khanikaev}},\ }\href {\doibase
  10.1038/nphoton.2016.253} {\bibfield  {journal} {\bibinfo  {journal} {Nat.
  Photonics}\ }\textbf {\bibinfo {volume} {11}},\ \bibinfo {pages} {130}
  (\bibinfo {year} {2017})}\BibitemShut {NoStop}%
\bibitem [{\citenamefont {Kallos}, \citenamefont {Chremmos},\ and\
  \citenamefont {Yannopapas}(2012)}]{Kallos_PhysRevB_2012}%
  \BibitemOpen
  \bibfield  {author} {\bibinfo {author} {\bibfnamefont {E.}~\bibnamefont
  {Kallos}}, \bibinfo {author} {\bibfnamefont {I.}~\bibnamefont {Chremmos}}, \
  and\ \bibinfo {author} {\bibfnamefont {V.}~\bibnamefont {Yannopapas}},\
  }\href {\doibase 10.1103/PhysRevB.86.245108} {\bibfield  {journal} {\bibinfo
  {journal} {Phys. Rev. B}\ }\textbf {\bibinfo {volume} {86}},\ \bibinfo
  {pages} {245108} (\bibinfo {year} {2012})}\BibitemShut {NoStop}%
\bibitem [{\citenamefont {Altman}\ and\ \citenamefont
  {Suchy}(1991)}]{Altman_book_1991}%
  \BibitemOpen
  \bibfield  {author} {\bibinfo {author} {\bibfnamefont {C.}~\bibnamefont
  {Altman}}\ and\ \bibinfo {author} {\bibfnamefont {K.}~\bibnamefont {Suchy}},\
  }\href@noop {} {\emph {\bibinfo {title} {Reciprocity, Spatial Mapping and
  Time Reversal in Electromagnetics}}}\ (\bibinfo  {publisher} {Kluwer,
  Dordrecht},\ \bibinfo {year} {1991})\BibitemShut {NoStop}%
\bibitem [{\citenamefont {Barybin}\ and\ \citenamefont
  {Dmitriev}(2002)}]{dmitriev_book_2002}%
  \BibitemOpen
  \bibfield  {author} {\bibinfo {author} {\bibfnamefont {A.~A.}\ \bibnamefont
  {Barybin}}\ and\ \bibinfo {author} {\bibfnamefont {V.~A.}\ \bibnamefont
  {Dmitriev}},\ }\href@noop {} {\emph {\bibinfo {title} {Modern Electrodynamics
  and Coupled-Mode Theory: Application to Guided-Wave Optics}}}\ (\bibinfo
  {publisher} {Rinton Press Princeton, New Jersey},\ \bibinfo {year}
  {2002})\BibitemShut {NoStop}%
\bibitem [{\citenamefont {Padilla}(2007)}]{Padilla_OptExpress_2007}%
  \BibitemOpen
  \bibfield  {author} {\bibinfo {author} {\bibfnamefont {W.~J.}\ \bibnamefont
  {Padilla}},\ }\href {\doibase 10.1364/OE.15.001639} {\bibfield  {journal}
  {\bibinfo  {journal} {Opt. Express}\ }\textbf {\bibinfo {volume} {15}},\
  \bibinfo {pages} {1639} (\bibinfo {year} {2007})}\BibitemShut {NoStop}%
\bibitem [{\citenamefont {Prosvirnin}\ and\ \citenamefont
  {Zouhdi}(2003)}]{Zouhdi_Advances_2003}%
  \BibitemOpen
  \bibfield  {author} {\bibinfo {author} {\bibfnamefont {S.}~\bibnamefont
  {Prosvirnin}}\ and\ \bibinfo {author} {\bibfnamefont {S.}~\bibnamefont
  {Zouhdi}},\ }in\ \href@noop {} {\emph {\bibinfo {booktitle} {Advances in
  Electromagnetics of Complex Media and Metamaterials}}},\ \bibinfo {editor}
  {edited by\ \bibinfo {editor} {\bibfnamefont {S.}~\bibnamefont {Zouhdi}}\
  and\ \bibinfo {editor} {\bibfnamefont {M.}~\bibnamefont {Arsalane}}}\
  (\bibinfo  {publisher} {Kluwer Academic Publishers},\ \bibinfo {address} {the
  Netherlands},\ \bibinfo {year} {2003})\ pp.\ \bibinfo {pages}
  {281--290}\BibitemShut {NoStop}%
\bibitem [{\citenamefont {Fedotov}\ \emph {et~al.}(2007)\citenamefont
  {Fedotov}, \citenamefont {Rose}, \citenamefont {Prosvirnin}, \citenamefont
  {Papasimakis},\ and\ \citenamefont {Zheludev}}]{Fedotov_PhysRevLett_2007}%
  \BibitemOpen
  \bibfield  {author} {\bibinfo {author} {\bibfnamefont {V.~A.}\ \bibnamefont
  {Fedotov}}, \bibinfo {author} {\bibfnamefont {M.}~\bibnamefont {Rose}},
  \bibinfo {author} {\bibfnamefont {S.~L.}\ \bibnamefont {Prosvirnin}},
  \bibinfo {author} {\bibfnamefont {N.}~\bibnamefont {Papasimakis}}, \ and\
  \bibinfo {author} {\bibfnamefont {N.~I.}\ \bibnamefont {Zheludev}},\ }\href
  {\doibase 10.1103/PhysRevLett.99.147401} {\bibfield  {journal} {\bibinfo
  {journal} {Phys. Rev. Lett.}\ }\textbf {\bibinfo {volume} {99}},\ \bibinfo
  {pages} {147401} (\bibinfo {year} {2007})}\BibitemShut {NoStop}%
\bibitem [{\citenamefont {Khardikov}, \citenamefont {Iarko},\ and\
  \citenamefont {Prosvirnin}(2010)}]{khardikov_JOpt_2010}%
  \BibitemOpen
  \bibfield  {author} {\bibinfo {author} {\bibfnamefont {V.~V.}\ \bibnamefont
  {Khardikov}}, \bibinfo {author} {\bibfnamefont {E.~O.}\ \bibnamefont
  {Iarko}}, \ and\ \bibinfo {author} {\bibfnamefont {S.~L.}\ \bibnamefont
  {Prosvirnin}},\ }\href {\doibase 10.1088/2040-8978/12/4/045102} {\bibfield
  {journal} {\bibinfo  {journal} {J. Opt.}\ }\textbf {\bibinfo {volume} {12}},\
  \bibinfo {pages} {045102} (\bibinfo {year} {2010})}\BibitemShut {NoStop}%
\bibitem [{\citenamefont {Jain}\ \emph {et~al.}(2015)\citenamefont {Jain},
  \citenamefont {Moitra}, \citenamefont {Koschny}, \citenamefont {Valentine},\
  and\ \citenamefont {Soukoulis}}]{jain_advoptmater_2015}%
  \BibitemOpen
  \bibfield  {author} {\bibinfo {author} {\bibfnamefont {A.}~\bibnamefont
  {Jain}}, \bibinfo {author} {\bibfnamefont {P.}~\bibnamefont {Moitra}},
  \bibinfo {author} {\bibfnamefont {T.}~\bibnamefont {Koschny}}, \bibinfo
  {author} {\bibfnamefont {J.}~\bibnamefont {Valentine}}, \ and\ \bibinfo
  {author} {\bibfnamefont {C.~M.}\ \bibnamefont {Soukoulis}},\ }\href {\doibase
  10.1002/adom.201500222} {\bibfield  {journal} {\bibinfo  {journal} {Adv. Opt.
  Mater.}\ }\textbf {\bibinfo {volume} {3}},\ \bibinfo {pages} {1431} (\bibinfo
  {year} {2015})}\BibitemShut {NoStop}%
\bibitem [{\citenamefont {Tuz}, \citenamefont {Prosvirnin},\ and\ \citenamefont
  {Kochetova}(2010)}]{tuz_PhysRevB_2010}%
  \BibitemOpen
  \bibfield  {author} {\bibinfo {author} {\bibfnamefont {V.~R.}\ \bibnamefont
  {Tuz}}, \bibinfo {author} {\bibfnamefont {S.~L.}\ \bibnamefont {Prosvirnin}},
  \ and\ \bibinfo {author} {\bibfnamefont {L.~A.}\ \bibnamefont {Kochetova}},\
  }\href {\doibase 10.1103/PhysRevB.82.233402} {\bibfield  {journal} {\bibinfo
  {journal} {Phys. Rev. B}\ }\textbf {\bibinfo {volume} {82}},\ \bibinfo
  {pages} {233402} (\bibinfo {year} {2010})}\BibitemShut {NoStop}%
\bibitem [{\citenamefont {Tuz}, \citenamefont {Butylkin},\ and\ \citenamefont
  {Prosvirnin}(2012)}]{Tuz_JOpt_2012}%
  \BibitemOpen
  \bibfield  {author} {\bibinfo {author} {\bibfnamefont {V.~R.}\ \bibnamefont
  {Tuz}}, \bibinfo {author} {\bibfnamefont {V.~S.}\ \bibnamefont {Butylkin}}, \
  and\ \bibinfo {author} {\bibfnamefont {S.~L.}\ \bibnamefont {Prosvirnin}},\
  }\href {http://stacks.iop.org/2040-8986/14/i=4/a=045102} {\bibfield
  {journal} {\bibinfo  {journal} {J. Opt.}\ }\textbf {\bibinfo {volume} {14}},\
  \bibinfo {pages} {045102} (\bibinfo {year} {2012})}\BibitemShut {NoStop}%
\bibitem [{\citenamefont {Tuz}\ and\ \citenamefont
  {Prosvirnin}(2011)}]{Tuz_EurPhys_2011}%
  \BibitemOpen
  \bibfield  {author} {\bibinfo {author} {\bibfnamefont {V.~R.}\ \bibnamefont
  {Tuz}}\ and\ \bibinfo {author} {\bibfnamefont {S.~L.}\ \bibnamefont
  {Prosvirnin}},\ }\href {\doibase 10.1051/epjap/2011110145} {\bibfield
  {journal} {\bibinfo  {journal} {Eur. Phys. J. Appl. Phys.}\ }\textbf
  {\bibinfo {volume} {56}},\ \bibinfo {pages} {30401} (\bibinfo {year}
  {2011})}\BibitemShut {NoStop}%
\bibitem [{\citenamefont {Khardikov}\ \emph {et~al.}(2016)\citenamefont
  {Khardikov}, \citenamefont {Mladyonov}, \citenamefont {Prosvirnin},\ and\
  \citenamefont {Tuz}}]{Khardikov2016}%
  \BibitemOpen
  \bibfield  {author} {\bibinfo {author} {\bibfnamefont {V.}~\bibnamefont
  {Khardikov}}, \bibinfo {author} {\bibfnamefont {P.}~\bibnamefont
  {Mladyonov}}, \bibinfo {author} {\bibfnamefont {S.}~\bibnamefont
  {Prosvirnin}}, \ and\ \bibinfo {author} {\bibfnamefont {V.}~\bibnamefont
  {Tuz}},\ }in\ \href {\doibase 10.1007/978-94-017-7315-7_5} {\emph {\bibinfo
  {booktitle} {Contemporary Optoelectronics: Materials, Metamaterials and
  Device Applications}}},\ \bibinfo {editor} {edited by\ \bibinfo {editor}
  {\bibfnamefont {O.}~\bibnamefont {Shulika}}\ and\ \bibinfo {editor}
  {\bibfnamefont {I.}~\bibnamefont {Sukhoivanov}}}\ (\bibinfo  {publisher}
  {Springer Netherlands},\ \bibinfo {address} {Dordrecht},\ \bibinfo {year}
  {2016})\ Chap.~\bibinfo {chapter} {5}, pp.\ \bibinfo {pages}
  {81--98}\BibitemShut {NoStop}%
\bibitem [{\citenamefont {Tong}\ \emph {et~al.}(2016)\citenamefont {Tong},
  \citenamefont {Gong}, \citenamefont {Liu}, \citenamefont {Yuan},
  \citenamefont {Huang}, \citenamefont {Xia},\ and\ \citenamefont
  {Wang}}]{Tong_OptExpress_2016}%
  \BibitemOpen
  \bibfield  {author} {\bibinfo {author} {\bibfnamefont {W.}~\bibnamefont
  {Tong}}, \bibinfo {author} {\bibfnamefont {C.}~\bibnamefont {Gong}}, \bibinfo
  {author} {\bibfnamefont {X.}~\bibnamefont {Liu}}, \bibinfo {author}
  {\bibfnamefont {S.}~\bibnamefont {Yuan}}, \bibinfo {author} {\bibfnamefont
  {Q.}~\bibnamefont {Huang}}, \bibinfo {author} {\bibfnamefont
  {J.}~\bibnamefont {Xia}}, \ and\ \bibinfo {author} {\bibfnamefont
  {Y.}~\bibnamefont {Wang}},\ }\href {\doibase 10.1364/OE.24.019661} {\bibfield
   {journal} {\bibinfo  {journal} {Opt. Express}\ }\textbf {\bibinfo {volume}
  {24}},\ \bibinfo {pages} {19661} (\bibinfo {year} {2016})}\BibitemShut
  {NoStop}%
\bibitem [{\citenamefont {Cui}\ \emph {et~al.}(2018)\citenamefont {Cui},
  \citenamefont {Zhou}, \citenamefont {Yuan}, \citenamefont {Qiu},
  \citenamefont {Zhu}, \citenamefont {Wang}, \citenamefont {Li}, \citenamefont
  {Song}, \citenamefont {Huang}, \citenamefont {Wang}, \citenamefont {Zeng},\
  and\ \citenamefont {Xia}}]{Cui_acsphotonics_2018}%
  \BibitemOpen
  \bibfield  {author} {\bibinfo {author} {\bibfnamefont {C.}~\bibnamefont
  {Cui}}, \bibinfo {author} {\bibfnamefont {C.}~\bibnamefont {Zhou}}, \bibinfo
  {author} {\bibfnamefont {S.}~\bibnamefont {Yuan}}, \bibinfo {author}
  {\bibfnamefont {X.}~\bibnamefont {Qiu}}, \bibinfo {author} {\bibfnamefont
  {L.}~\bibnamefont {Zhu}}, \bibinfo {author} {\bibfnamefont {Y.}~\bibnamefont
  {Wang}}, \bibinfo {author} {\bibfnamefont {Y.}~\bibnamefont {Li}}, \bibinfo
  {author} {\bibfnamefont {J.}~\bibnamefont {Song}}, \bibinfo {author}
  {\bibfnamefont {Q.}~\bibnamefont {Huang}}, \bibinfo {author} {\bibfnamefont
  {Y.}~\bibnamefont {Wang}}, \bibinfo {author} {\bibfnamefont {C.}~\bibnamefont
  {Zeng}}, \ and\ \bibinfo {author} {\bibfnamefont {J.}~\bibnamefont {Xia}},\
  }\href {\doibase 10.1021/acsphotonics.8b00754} {\bibfield  {journal}
  {\bibinfo  {journal} {ACS Photonics}\ }\textbf {\bibinfo {volume} {5}},\
  \bibinfo {pages} {4074} (\bibinfo {year} {2018})}\BibitemShut {NoStop}%
\bibitem [{\citenamefont {Khardikov}, \citenamefont {Iarko},\ and\
  \citenamefont {Prosvirnin}(2012)}]{Khardikov_JOpt_2012}%
  \BibitemOpen
  \bibfield  {author} {\bibinfo {author} {\bibfnamefont {V.~V.}\ \bibnamefont
  {Khardikov}}, \bibinfo {author} {\bibfnamefont {E.~O.}\ \bibnamefont
  {Iarko}}, \ and\ \bibinfo {author} {\bibfnamefont {S.~L.}\ \bibnamefont
  {Prosvirnin}},\ }\href {http://stacks.iop.org/2040-8986/14/i=3/a=035103}
  {\bibfield  {journal} {\bibinfo  {journal} {J. Opt.}\ }\textbf {\bibinfo
  {volume} {14}},\ \bibinfo {pages} {035103} (\bibinfo {year}
  {2012})}\BibitemShut {NoStop}%
\bibitem [{\citenamefont {Zhang}, \citenamefont {MacDonald},\ and\
  \citenamefont {Zheludev}(2013)}]{Zhang_OptExpress_2013}%
  \BibitemOpen
  \bibfield  {author} {\bibinfo {author} {\bibfnamefont {J.}~\bibnamefont
  {Zhang}}, \bibinfo {author} {\bibfnamefont {K.~F.}\ \bibnamefont
  {MacDonald}}, \ and\ \bibinfo {author} {\bibfnamefont {N.~I.}\ \bibnamefont
  {Zheludev}},\ }\href {\doibase 10.1364/OE.21.026721} {\bibfield  {journal}
  {\bibinfo  {journal} {Opt. Express}\ }\textbf {\bibinfo {volume} {21}},\
  \bibinfo {pages} {26721} (\bibinfo {year} {2013})}\BibitemShut {NoStop}%
\bibitem [{\citenamefont {Tuz}\ \emph {et~al.}(2018)\citenamefont {Tuz},
  \citenamefont {Khardikov}, \citenamefont {Kupriianov}, \citenamefont
  {Domina}, \citenamefont {Xu}, \citenamefont {Wang},\ and\ \citenamefont
  {Sun}}]{Tuz_OptExpress_2018}%
  \BibitemOpen
  \bibfield  {author} {\bibinfo {author} {\bibfnamefont {V.~R.}\ \bibnamefont
  {Tuz}}, \bibinfo {author} {\bibfnamefont {V.~V.}\ \bibnamefont {Khardikov}},
  \bibinfo {author} {\bibfnamefont {A.~S.}\ \bibnamefont {Kupriianov}},
  \bibinfo {author} {\bibfnamefont {K.~L.}\ \bibnamefont {Domina}}, \bibinfo
  {author} {\bibfnamefont {S.}~\bibnamefont {Xu}}, \bibinfo {author}
  {\bibfnamefont {H.}~\bibnamefont {Wang}}, \ and\ \bibinfo {author}
  {\bibfnamefont {H.-B.}\ \bibnamefont {Sun}},\ }\href {\doibase
  10.1364/OE.26.002905} {\bibfield  {journal} {\bibinfo  {journal} {Opt.
  Express}\ }\textbf {\bibinfo {volume} {26}},\ \bibinfo {pages} {2905}
  (\bibinfo {year} {2018})}\BibitemShut {NoStop}%
\bibitem [{\citenamefont {Sayanskiy}\ \emph
  {et~al.}(2018{\natexlab{b}})\citenamefont {Sayanskiy}, \citenamefont
  {Kupriianov}, \citenamefont {Xu}, \citenamefont {Kapitanova}, \citenamefont
  {Dmitriev}, \citenamefont {Khardikov},\ and\ \citenamefont
  {Tuz}}]{sayanskiy_arXiv_2018}%
  \BibitemOpen
  \bibfield  {author} {\bibinfo {author} {\bibfnamefont {A.}~\bibnamefont
  {Sayanskiy}}, \bibinfo {author} {\bibfnamefont {A.~S.}\ \bibnamefont
  {Kupriianov}}, \bibinfo {author} {\bibfnamefont {S.}~\bibnamefont {Xu}},
  \bibinfo {author} {\bibfnamefont {P.}~\bibnamefont {Kapitanova}}, \bibinfo
  {author} {\bibfnamefont {V.}~\bibnamefont {Dmitriev}}, \bibinfo {author}
  {\bibfnamefont {V.~V.}\ \bibnamefont {Khardikov}}, \ and\ \bibinfo {author}
  {\bibfnamefont {V.~R.}\ \bibnamefont {Tuz}},\ }\href@noop {} {\bibfield
  {journal} {\bibinfo  {journal} {arXiv preprint arXiv:1811.11396}\ } (\bibinfo
  {year} {2018}{\natexlab{b}})}\BibitemShut {NoStop}%
\bibitem [{\citenamefont {Dmitriev}(2011)}]{Dmitriev_Metamat_2011}%
  \BibitemOpen
  \bibfield  {author} {\bibinfo {author} {\bibfnamefont {V.}~\bibnamefont
  {Dmitriev}},\ }\href {\doibase https://doi.org/10.1016/j.metmat.2011.04.003}
  {\bibfield  {journal} {\bibinfo  {journal} {Metamaterials}\ }\textbf
  {\bibinfo {volume} {5}},\ \bibinfo {pages} {14} (\bibinfo {year}
  {2011})}\BibitemShut {NoStop}%
\bibitem [{\citenamefont {Dmitriev}(2013)}]{Dmitriev_IEEEAntennas_2013}%
  \BibitemOpen
  \bibfield  {author} {\bibinfo {author} {\bibfnamefont {V.}~\bibnamefont
  {Dmitriev}},\ }\href {\doibase 10.1109/TAP.2012.2220316} {\bibfield
  {journal} {\bibinfo  {journal} {IEEE Trans. Antennas Propag.}\ }\textbf
  {\bibinfo {volume} {61}},\ \bibinfo {pages} {185} (\bibinfo {year}
  {2013})}\BibitemShut {NoStop}%
\bibitem [{\citenamefont {Hamermesh}\ and\ \citenamefont
  {Mullin}(1962)}]{hamermesh_book_1962}%
  \BibitemOpen
  \bibfield  {author} {\bibinfo {author} {\bibfnamefont {M.}~\bibnamefont
  {Hamermesh}}\ and\ \bibinfo {author} {\bibfnamefont {A.~A.}\ \bibnamefont
  {Mullin}},\ }\href@noop {} {\emph {\bibinfo {title} {Group Theory and its
  Applications to Physical Problems}}}\ (\bibinfo  {publisher} {Dover
  Publictions Inc., New York},\ \bibinfo {year} {1962})\BibitemShut {NoStop}%
\bibitem [{com()}]{comsol}%
  \BibitemOpen
  \href
  {https://www.comsol.com/model/frequency-selective-surface-periodic-complementary-split-ring-resonator-15711}
  {\enquote {\bibinfo {title} {Frequency selective surface, periodic
  complementary split ring resonator},}\ }\bibinfo {note} {Comsol Application
  Gallery {ID:~15711}}\BibitemShut {NoStop}%
\bibitem [{\citenamefont {Decker}\ \emph {et~al.}(2009)\citenamefont {Decker},
  \citenamefont {Burger}, \citenamefont {Linden},\ and\ \citenamefont
  {Wegener}}]{Wegener_PhysRevB_2009}%
  \BibitemOpen
  \bibfield  {author} {\bibinfo {author} {\bibfnamefont {M.}~\bibnamefont
  {Decker}}, \bibinfo {author} {\bibfnamefont {S.}~\bibnamefont {Burger}},
  \bibinfo {author} {\bibfnamefont {S.}~\bibnamefont {Linden}}, \ and\ \bibinfo
  {author} {\bibfnamefont {M.}~\bibnamefont {Wegener}},\ }\href {\doibase
  10.1103/PhysRevB.80.193102} {\bibfield  {journal} {\bibinfo  {journal} {Phys.
  Rev. B}\ }\textbf {\bibinfo {volume} {80}},\ \bibinfo {pages} {193102}
  (\bibinfo {year} {2009})}\BibitemShut {NoStop}%
\bibitem [{\citenamefont {Decker}, \citenamefont {Linden},\ and\ \citenamefont
  {Wegener}(2009)}]{Decker_OptLett_2009}%
  \BibitemOpen
  \bibfield  {author} {\bibinfo {author} {\bibfnamefont {M.}~\bibnamefont
  {Decker}}, \bibinfo {author} {\bibfnamefont {S.}~\bibnamefont {Linden}}, \
  and\ \bibinfo {author} {\bibfnamefont {M.}~\bibnamefont {Wegener}},\ }\href
  {\doibase 10.1364/OL.34.001579} {\bibfield  {journal} {\bibinfo  {journal}
  {Opt. Lett.}\ }\textbf {\bibinfo {volume} {34}},\ \bibinfo {pages} {1579}
  (\bibinfo {year} {2009})}\BibitemShut {NoStop}%
\bibitem [{\citenamefont {Landau}\ and\ \citenamefont
  {Lifshitz}(1980)}]{landau_1960_5}%
  \BibitemOpen
  \bibfield  {author} {\bibinfo {author} {\bibfnamefont {L.~D.}\ \bibnamefont
  {Landau}}\ and\ \bibinfo {author} {\bibfnamefont {E.~M.}\ \bibnamefont
  {Lifshitz}},\ }\href@noop {} {\emph {\bibinfo {title} {Course of Theoretical
  Physics: Statistical Physics}}},\ \bibinfo {edition} {3rd}\ ed.,\
  Vol.~\bibinfo {volume} {5}\ (\bibinfo  {publisher} {Butterworth-Heinemann,
  Oxford},\ \bibinfo {year} {1980})\ Chap.~\bibinfo {chapter} {14}\BibitemShut
  {NoStop}%
\bibitem [{\citenamefont {Cong}\ \emph {et~al.}(2018)\citenamefont {Cong},
  \citenamefont {Savinov}, \citenamefont {Srivastava}, \citenamefont {Han},\
  and\ \citenamefont {Singh}}]{Singh_AdvMat_2018}%
  \BibitemOpen
  \bibfield  {author} {\bibinfo {author} {\bibfnamefont {L.}~\bibnamefont
  {Cong}}, \bibinfo {author} {\bibfnamefont {V.}~\bibnamefont {Savinov}},
  \bibinfo {author} {\bibfnamefont {Y.~K.}\ \bibnamefont {Srivastava}},
  \bibinfo {author} {\bibfnamefont {S.}~\bibnamefont {Han}}, \ and\ \bibinfo
  {author} {\bibfnamefont {R.}~\bibnamefont {Singh}},\ }\href {\doibase
  10.1002/adma.201804210} {\bibfield  {journal} {\bibinfo  {journal} {Adv.
  Mater.}\ }\textbf {\bibinfo {volume} {30}},\ \bibinfo {pages} {1804210}
  (\bibinfo {year} {2018})}\BibitemShut {NoStop}%
\bibitem [{\citenamefont {Jackson}(1998)}]{Jackson_1998}%
  \BibitemOpen
  \bibfield  {author} {\bibinfo {author} {\bibfnamefont {J.~D.}\ \bibnamefont
  {Jackson}},\ }\href@noop {} {\emph {\bibinfo {title} {Classical
  Electrodynamics}}},\ \bibinfo {edition} {3rd}\ ed.\ (\bibinfo  {publisher}
  {Wiley},\ \bibinfo {year} {1998})\BibitemShut {NoStop}%
\bibitem [{\citenamefont {Baryshnikova}\ \emph {et~al.}(2017)\citenamefont
  {Baryshnikova}, \citenamefont {Novitsky}, \citenamefont {Evlyukhin},\ and\
  \citenamefont {Shalin}}]{Baryshnikova_JOSA_B_2017}%
  \BibitemOpen
  \bibfield  {author} {\bibinfo {author} {\bibfnamefont {K.~V.}\ \bibnamefont
  {Baryshnikova}}, \bibinfo {author} {\bibfnamefont {A.}~\bibnamefont
  {Novitsky}}, \bibinfo {author} {\bibfnamefont {A.~B.}\ \bibnamefont
  {Evlyukhin}}, \ and\ \bibinfo {author} {\bibfnamefont {A.~S.}\ \bibnamefont
  {Shalin}},\ }\href {\doibase 10.1364/JOSAB.34.000D36} {\bibfield  {journal}
  {\bibinfo  {journal} {J. Opt. Soc. Am. B}\ }\textbf {\bibinfo {volume}
  {34}},\ \bibinfo {pages} {D36} (\bibinfo {year} {2017})}\BibitemShut
  {NoStop}%
\end{thebibliography}%

\end{document}